\pdfoutput=1
\documentclass[fleqn,usenatbib]{mnras}

\usepackage{newtxtext,newtxmath}

\usepackage[T1]{fontenc}
\usepackage{ae,aecompl}
\usepackage{ulem}
\usepackage{soul}
\usepackage{color}
\hypersetup{
    colorlinks=true,
    linkcolor=blue,
    filecolor=magenta,      
    urlcolor=blue,
}

\newcommand{\msun}{\mbox{$M_{\odot}$}}

\def\vector#1{\mbox{\boldmath $#1$}}

\newcommand{\pc}{\ensuremath{\,\mathrm{pc}}}
\newcommand{\kpc}{\ensuremath{\,\mathrm{kpc}}}

\newcommand{\Gyr}{\ensuremath{\,\mathrm{Gyr}}}
\newcommand{\kms}{\ensuremath{\,\mathrm{km\ s}^{-1}}}

\newcommand{\masyr}{\ensuremath{\,\mathrm{mas\ yr}^{-1}}}

\newcommand{\DM}{\mathcal{D}}
\newcommand{\vlos}{v_{\ensuremath{\mathrm{los}}}}
\newcommand{\muell}{\mu_{\ell*}}
\newcommand{\mub}{\mu_{b}}
\newcommand{\mualpha}{\mu_{\alpha*}}
\newcommand{\mudelta}{\mu_\delta}

\newcommand{\flag}[1]{\texttt{\lowercase{#1}}}
\newcommand{\agama}{\texttt{AGAMA}}
\newcommand{\emcee}{\flag{emcee}}
\newcommand{\Gaia}{\textit{Gaia}}

\newcommand{\eq}[1]{\begin{align}#1\end{align}}

\usepackage{graphicx}	
\usepackage{amsmath}	
\usepackage{mathrsfs}
\usepackage{color}

\title[Distribution function fitting]
{Action-based distribution function modelling for constraining the shape of the Galactic dark matter halo}

\author[K. Hattori et al.]{
Kohei Hattori,$^{1,2,3,4,5}$\thanks{E-mail: khattori@ism.ac.jp}
Monica Valluri,$^{5}$
Eugene Vasiliev$^{6}$
\\
$^{1}$
National Astronomical Observatory of Japan  
2-21-1 Osawa, Mitaka, Tokyo 181-0015, Japan\\
$^{2}$
Institute of Statistical Mathematics
10-3 Midoricho, Tachikawa, Tokyo 190-0014, Japan\\
$^{3}$
McWilliams Center for Cosmology, Carnegie Mellon University 
5000 Forbes Avenue, Pittsburgh, PA 15213, USA\\
$^{4}$
Department of Physics, Carnegie Mellon University,
5000 Forbes Avenue, Pittsburgh, PA 15213, USA\\
$^{5}$
Department of Astronomy, University of Michigan,
1085 S.\ University Avenue, Ann Arbor, MI 48109, USA\\
$^{6}$Institute of Astronomy, University of Cambridge, Madingley Rd, Cambridge CB3 0HA, UK\\
}

\date{Accepted XXX. Received YYY; in original form ZZZ}

\pubyear{2020}

\begin{document}
\label{firstpage}
\pagerange{\pageref{firstpage}--\pageref{lastpage}}
\maketitle

\begin{abstract}

We estimate the 3D density profile of the Galactic dark matter (DM) halo within $r \lesssim 30 \;\mathrm{kpc}$ from the Galactic centre 
by using the astrometric data for halo RR Lyrae stars from Gaia DR2. We model both  the stellar halo distribution function and the Galactic potential, fully taking into account the survey selection function, the observational errors, and the missing line-of-sight velocity data for RR Lyrae stars. 
With a Bayesian MCMC method, we infer the model parameters, including the density flattening of the DM halo $q$, which is assumed to be constant as a function of radius. We find that 99\% of the posterior distribution of $q$ is located at $q>0.963$, which strongly disfavours a flattened DM halo. We cannot draw any conclusions as to whether the Galactic DM halo at $r \lesssim 30 \kpc$ is prolate, because we restrict ourselves to axisymmetric oblate halo models with $q\leq1$.
Our result is inconsistent with predictions from cosmological hydrodynamical simulations that advocate more oblate ($\langle{q}\rangle \sim0.8 \pm 0.15$) DM halos within $\sim 15\%$ of the virial radius for Milky-Way-sized galaxies. An alternative possibility, based on our validation tests with a cosmological simulation, is that the true value $q$ of the Galactic halo could be consistent with cosmological simulations but that disequilibrium in the Milky Way potential is inflating our measurement of $q$ by 0.1-0.2.
As a by-product of our analysis, 
our model constrains the DM density in the Solar neighbourhood to be
$\rho_{\mathrm{DM},\odot} = (9.01^{+0.18}_{-0.20})\times10^{-3}M_\odot \mathrm{pc}^{-3} = 0.342^{+0.007}_{-0.007}$ 
$\;\mathrm{GeV} \mathrm{cm}^{-3}$.

\end{abstract}

\begin{keywords}
   stars: variables: RR Lyrae 
-- Galaxy: halo 
-- Galaxy: kinematics and dynamics
-- Galaxy: structure
\end{keywords}



\section{Introduction}
\label{sec:intro}

Determining the mass distribution of the dark matter (DM) halo of the Milky Way (MW) is currently of the most important tasks in Galactic astronomy in the cosmological context,because the shape of the Galactic DM halo can be used to test the prediction from $\Lambda$CDM cosmology. 

There have been numerous efforts to measure the density profiles and shapes of DM haloes arising from cosmological N-body simulations. DM-only simulations show that the spherically averaged radial density profiles of DM halos follow a universal form \citep{NFW1997} and that their 3-dimensional (3D) mass distributions are triaxial \citep{JingSuto2002}. It has also been recognised that the DM halo's shape is affected by the infall and condensation of baryons  \citep{Dubinski1994, Kazantzidis2004, Kazantzidis2010, Zemp2012, Zhu2016}. This result has recently been confirmed on 10,000 halos with masses in the range $10^{11}-3\times10^{14}\msun$ from the Illustris suite of simulations with and without baryonic physics \citep{Chua2019}.  In particular, DM haloes of  MW-sized galaxies with baryons have a nearly oblate axisymmetric shape, with a minor-to-major axis ratio of $0.79 \pm 0.15$ especially within  $0.15 r_{200}$. 
(Here, $r_{200}$ is the virial radius within which the average density is 200 times the critical density of the Universe.)
In contrast,  DM-only simulations in the same mass range have DM haloes that are highly triaxial. A similar trend is also reported by \cite{Prada2019}, 
who used MW-sized haloes in Auriga simulations.  
The deformation of the DM haloes by the baryons has been shown to arise due to the deformation in the shapes of the orbits of DM particles by the more concentrated baryonic components in both controlled and cosmological simulations \citep{Debattista_etal_2008,Valluri_etal_2010, Valluri_etal_2012}. 

In summary  cosmological hydrodynamical simulations over the past two decades have consistently predicted that the DM haloes of MW-sized galaxies are nearly oblate-axisymmetric especially in the inner regions ($<30$-$50 \kpc$), becoming more triaxial at larger radii.

In this paper we test this prediction  by measuring the mass distribution of the Galactic DM halo using halo stars as kinematic tracers of the 3D mass distribution of the MW halo.

There are two broad categories of methods that have frequently been used to measure the shape of the MW's DM halo.\footnote{
While other methods to measure the DM halo's shape have been proposed, they have not been used very extensively. 
For example, \cite{Olling2000} used the flaring of the $\mathrm{H_I}$ gas disc to determine the flattening of the DM halo, and \cite{Gnedin2005} proposed the use of proper motions of hyper-velocity stars to derive the triaxiality of the halo. 
}
The first  uses the kinematics and spatial distribution of stellar streams in the halo
\citep{Johnston1999, Helmi2004, Johnston2005, Fellhauer2006, Koposov2010, Law2010, Sanders2013, Bovy2014, Gibbons2014, Bowden2015, Kupper2015, Bovy2016, Malhan2019, Vasiliev2020Tango}. 
This method relies on the fact that stellar streams are remnants of tidally disrupted dwarf galaxies or star clusters,  and that stars that form the stellar stream have  coherent motions and show a 1-dimensional spatial distribution that approximately traces the orbit of the progenitor system.

A second method is to use the kinematics of 
halo tracers such as globular clusters or halo field stars 
\citep{Loebman2014, Bowden2016, Cole2017, Eadie2019, Posti2019, Watkins2019, Wegg2019,Nitschai2020}. 
In this approach, the halo tracers are assumed to be in dynamical equilibrium, and the task is to determine the Galactic potential such that the observed position and velocity of halo tracers 
are consistent with this equilibrium assumption. 
Classically, most of the studies in this avenue 
used the axisymmetric form of the Jeans equations (e.g., \citealt{Wegg2019}). Following the availability of accurate astrometric data from Gaia DR2 \citep{Lindegren2018}, there have been some attempts to model the positions and velocities of the halo tracers  with the phase-space distribution function (DF) \citep{McMillanBinney2012, McMillanBinney2013, Binney2015, Cole2017, Posti2019}. 
Our work is a continuation of these works, 
and we use DF models to infer the Galactic potential.

In this paper, we infer the 3D mass distribution of the Galactic DM halo by simultaneously modelling the DF of the halo stars and the MW potential.
The `DF fitting' technique was pioneered by \cite{McMillanBinney2012,McMillanBinney2013} and extended by \cite{Ting2013} and \cite{Trick2016}. The most important improvement of our implementation of this method is a new way of handling the distance errors on the sample stars, which was neglected in previous works \citep{McMillanBinney2012,McMillanBinney2013}.

In previous works of a similar nature, \cite{Das2016a} and \cite{Das2016b} modelled the stellar halo DF (and its metallicity dependence) while keeping the MW potential fixed. 
Also, \cite{Binney2015} and \cite{Cole2017} modelled the DF of the DM halo, while we only model the spatial density distribution of the DM halo.

The outline of this paper is as follows. 
In Section \ref{sec:data}, 
we describe the data sets used in this paper. 
In Section \ref{sec:model}, 
we describe our model ingredients. 
In Section \ref{sec:DFlikelihood},
we describe how we evaluate the likelihood function from the stellar halo data. 
In Section \ref{sec:analysis},
we describe the setup for our Bayesian analysis. 
In Section \ref{sec:result},
we present our results. 
In Section \ref{sec:discussion}, 
we compare our results with the literature in this field
and we mention some issues in our analysis. 
In Section \ref{sec:conclusion}, 
we present our conclusion. 

In Appendix  \ref{sec:validation} we present validation tests of our method on a variety of mock stellar halo data.

\section{Data} \label{sec:data}

Here we describe three sets of kinematic data used in this paper. We note that the coordinate system is defined in Appendix \ref{appendix:coordinate}.

\subsection{Circular velocity}
\label{sec:vcirc}

The best way of constraining the Galactic potential in the disc plane is to use the circular velocity $v_\mathrm{circ}(R)$.  In our analysis, we use the measurement of $v_\mathrm{circ}(R)$ between $5 \leq R \leq 25\kpc$ from  \citet{Eilers2019}. This measurement was obtained by applying Jeans analysis to 6-dimensional (6D) phase-space coordinates for more than 23000 disc red giants  with photometric data from 2MASS, WISE and Gaia DR2 for which precise spectro-photometric distances were derived from APOGEE DR14 \citep{Hogg2019}.

Due to accurate distances, the uncertainty in $v_\mathrm{circ}$ is small in \cite{Eilers2019}.  The random error is $\sigma_\mathrm{circ,rand} \simeq$ (1-3) $\kms$ at $R \lesssim 20 \kpc$,   and $\sigma_\mathrm{circ,rand} \sim 10 \kms$ at $20 \kpc \leq R \leq 25 \kpc$. The systematic error is roughly 
$\sigma_\mathrm{circ,sys} \simeq$ (0.02-0.05) $\times v_\mathrm{circ}$. The quoted systematic error arises from various assumptions in the way they performed the Jeans analysis, such as the assumed density profile of the stellar disc. We do not take into account the systematic error in our main analysis, but we have verified that the inclusion of the systematic error hardly changes our main conclusions.

\subsection{Vertical force $K_z$ at $z=1.1$ kpc}

Classically, the vertical component of the force at $z=1.1 \kpc$ away from the disc plane, $K_{z, 1.1 \kpc} (R)$, has been measured via the vertical Jeans equation \citep{KuijkenGilmore1991}. We use a recent measurement of $K_{z, 1.1 \kpc}(R)$ by \cite{BovyRix2013} to aid our measurement of the gravitational potential. To be specific, we used the values of $R-R_0$ and $K_{z, 1.1 \kpc}(R)$ in their Table 3, by adopting $R_0 = 8.178 \kpc$.

We note that \cite{BovyRix2013} derived $K_{z, 1.1 \kpc}(R)$ by analysing the DFs 
of various mono-abundance  populations in the thin and thick discs of the Milky Way from SDSS/SEGUE data. Thus, their data are independent of the halo RR Lyrae  data 
discussed in Section \ref{sec:RRLdata}. Also, while the circular velocity data (discussed in Section~\ref{sec:vcirc}) constraints the radial force in the Galactic disc plane, $K_{z, 1.1 \kpc}(R)$ constraints the vertical force away from the disc plane. 
In this regard, $v_\mathrm{circ}$ and the $K_{z, 1.1 \kpc}$ are complementary data.

\subsection{The RR Lyrae sample: Five-dimensional phase-space coordinates} 
\label{sec:RRLdata}

\subsubsection{Sample selection} \label{sec:selection}

To measure the global 3D shape of the Galactic potential, we need kinematic data for halo tracers. 
We use the `clean' sample of 93,345 RR Lyrae stars 
compiled by \cite{Iorio2019} (and kindly provided by Dr. Giuliano Iorio), which were selected from the Gaia DR2 catalogue. This sample does not include  stars associated with known halo substructure (such as the core of the Sagittarius stream, satellite galaxies, or globular clusters).
\footnote{Since the fraction of excluded stars is negligible, it is safe to regard our sample as a kinematically unbiased sample.}

In this paper, we focus on the 3D shape of the DM halo within $r \lesssim 30 \kpc$, where cosmological simulations suggest that the dark halo's shape is more or less oblate-axisymmetric due to the influence from the baryonic potential \citep[e.g.][]{Kazantzidis2004,Zemp2012,Chua2019}.
We apply a simple spatial cut to the `clean' RR Lyrae sample to extract inner halo stars. To be specific, 
for each star in the sample, we first use the observed photometric distance $d$ (neglecting the distance uncertainty) to 
determine the 3D Galactocentric coordinate $(x,y,z)$. 
Then we select those stars that satisfy 
\eq{\label{eq:obs_selection}
|b| \geq 20^\circ, \quad 
|z| \geq 5 \kpc, \quad
d \leq 20 \kpc,
}
where $b$ is the Galactic latitude. 
After these cuts, we are left with $N_\mathrm{RRL}=16197$ RR Lyrae stars. This simple spatial selection has a number of advantages. First, we can safely exclude RR Lyrae stars associated with the stellar disc. Secondly, we can avoid using RR Lyrae stars located in low-completeness regions, 
such as regions that are highly dust-obscured  or have a high degree of crowding \citep{Iorio2019}. Lastly, by using a well-defined spatial selection function, our analysis 
becomes mathematically tractable (see Section \ref{sec:DFlikelihood}).

\subsubsection{Content of the RR Lyrae sample} \label{sec:RRLcontent}

Following \cite{McMillanBinney2012}, we define a 6D observational vector for each RR Lyrae star
\eq{\label{eq:observable_vector}
\vector{u} = (\ell, b, \DM, \mualpha, \mudelta, \vlos),
}
where 
$(\ell, b)$ are the Galactic coordinates, $\DM$ is the distance modulus, $(\mu_{\alpha*}, \mu_\delta)$ are the proper motions in ICRS coordinates, and 
$\vlos$ is the line-of-sight velocity. 
We note that we will not use $\vlos$ in our main analysis, 
because $\vlos$ is not available for the majority of the RR Lyrae stars. As we shall see in later Sections, 
we will fit only 5-dimensional (5D) data for RR Lyrae stars and marginalise over the value of $\vlos$. 
(Mock analyses using 6D mock data are presented in Appendix \ref{sec:validation}.)

In this paper, we neglect the errors in $(\ell, b)$. 
In our RR Lyrae catalogue, the distance is estimated from period-luminosity relationship. The uncertainty in the distance modulus is quite uniform, with $\sigma_{\DM} = 0.2401 \pm 0.056$ (the fractional distance error is $11.07 \pm 0.26$\%). Thus, in our analysis, we assume that the errors in $\DM$ as $\sigma_{\DM} = 0.240$ for all the stars. This assumption of uniform error in $\DM$ plays 
an important role in simplifying our DF fitting. 


\begin{figure*}
\centering
\includegraphics[width=3.3in]{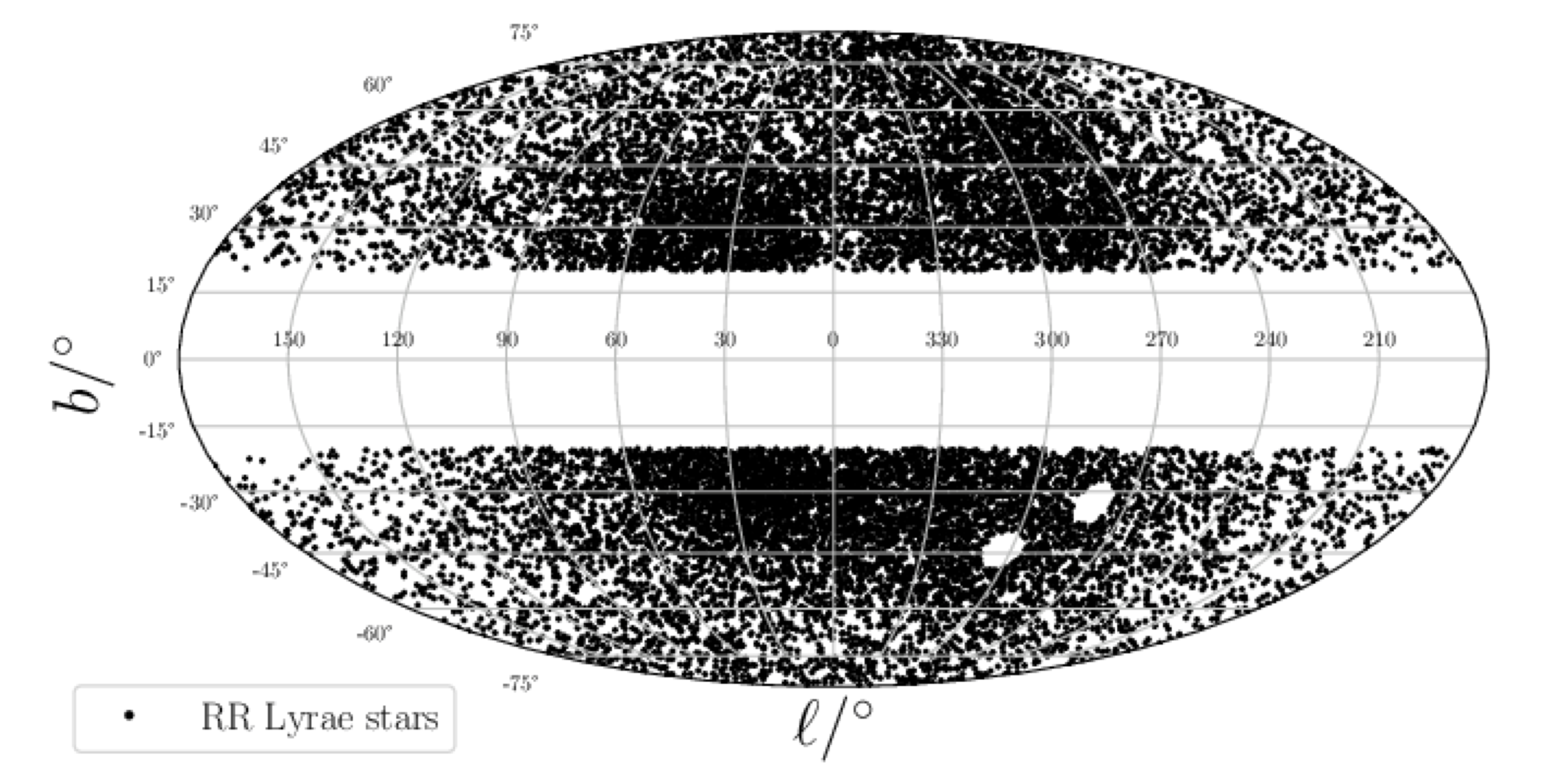}
\includegraphics[width=3.3in]{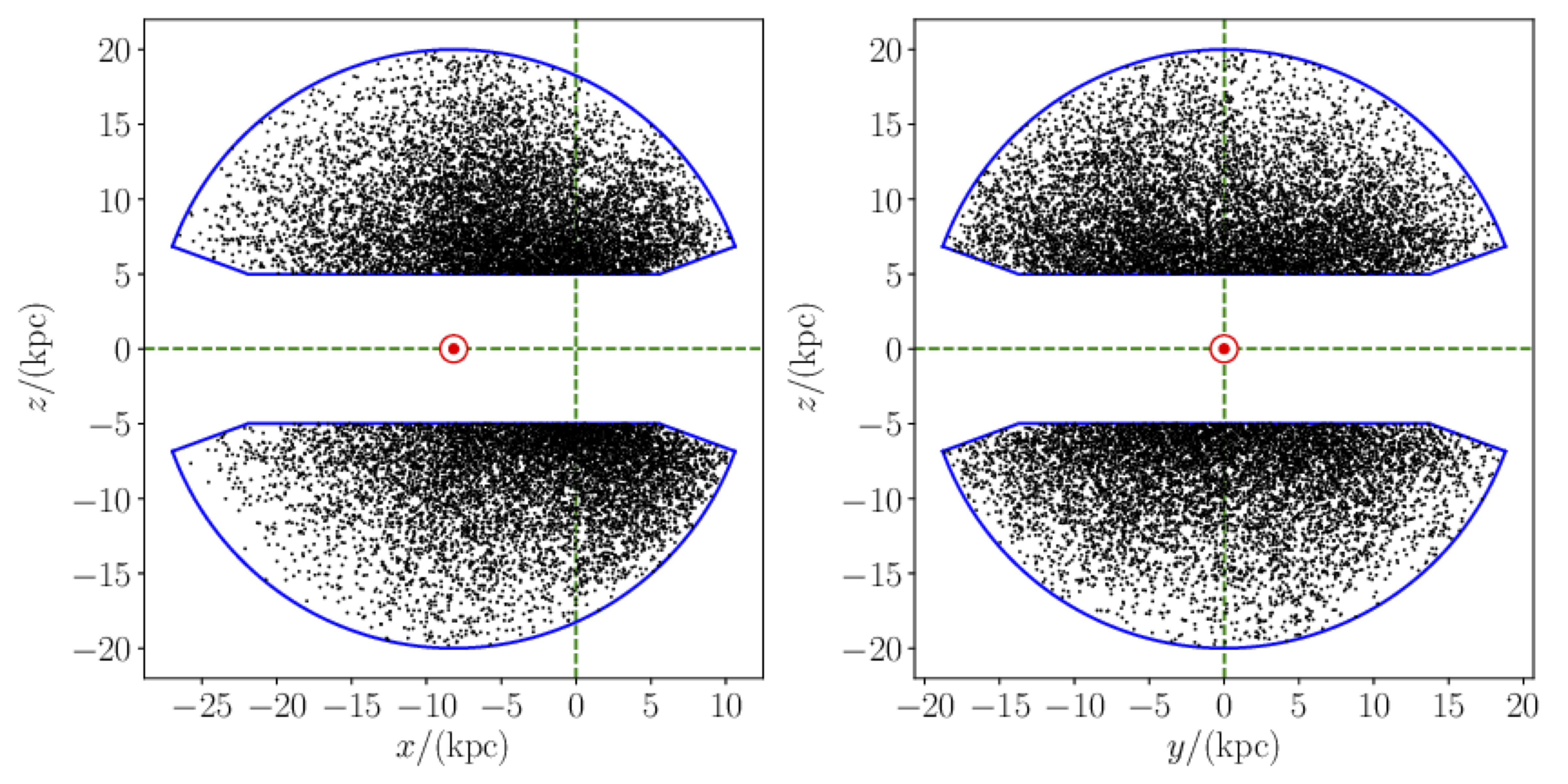}
\caption{
Distribution of the 16197 stars in our RR Lyrae sample 
on the sky ({\bf left}), in the Galactocentric Cartesian $(x,z)$-plane ({\bf middle}) and $(y,z)$-plane ({\bf right}). 
Our sample does not include stars associated with obvious substructure, such as the Large and Small Magellanic Clouds (two small holes in the left panel at around $270^\circ<\ell<315^\circ$, $-50^\circ<b<-25^\circ$). 
In the middle and right panels, the Solar location is marked with the red $\odot$ symbol. Our sample selection criteria only select stars that are confined within the region enclosed by the solid blue curve. 
}
\label{fig:RRLdistribution}
\end{figure*}

\section{Model} \label{sec:model}

Our goal is to estimate the density of the DM halo, 
$\rho_{\rm DM}(R,z)$. To achieve this goal, we use parametric models for the
(1) MW potential $\Phi(R,z)$ (consisting of baryonic and DM components); 
and 
(2) stellar halo DF $f(\vector{J})$ and estimate the parameters of these components with a Bayesian analysis.  Throughout this paper, we assume that the MW is static and axisymmetric, and that the stellar halo is in dynamical equilibrium. While this assumption has been recently called into question by modelling of  tidal streams \citep{erkal2019, Vasiliev2020Tango} and the kinematics of outer halo stars \citep{Petersen2020arXiv}, in this paper (as in most works that use distributed field halo stars as kinematic tracers) we ignore this disequilibrium. 
In this Section, we describe the ingredients of our equilibrium model.

\subsection{Models for the gravitational potential}

We model $\Phi(R,z)$ following the parametrisation by \cite{McMillan2017}, 
except that we allow the DM distribution to be oblate-axisymmetric rather than spherical. First, we briefly summarise the forms assumed for the baryonic  mass components
(the bulge, stellar discs and gas discs) which are identical to \cite{McMillan2017}. We then describe our assumed DM distribution model. 

\subsubsection{The bulge}

For the bulge we adopt a function of the form, 
\eq{
\rho_\mathrm{bulge}(R,z) =  
\frac{c_b \times M_\mathrm{bulge}}{(1 + (r'/0.075 \kpc))^{1.8}} 
\exp \left[ -\left( \frac{r'}{2.1 \kpc} \right)^{2} \right],  
}
where $(r')^2 = R^2 + (z/0.5)^2$. 
The total bulge mass is $M_\mathrm{bulge}$, and $c_b$ is a normalisation constant. $M_\mathrm{bulge}$ is treated as a free parameter, with a prior of 
$M_\mathrm{bulge}=(8.90\pm0.89)\times 10^9 M_\odot$ (see Table \ref{tab:maxL}).

\subsubsection{The thin and thick stellar discs} 

We adopt density profiles for the thin and thick discs of the form,
\eq{
\rho_{\operatorname{stellar-disc}}(R,z) = 
\frac{M_\mathrm{disc}}{4 \pi R_\mathrm{d}^2 z_\mathrm{d}} 
\exp \left[ - \frac{|z|}{z_\mathrm{d}} - \frac{R}{R_\mathrm{d}} \right], 
}
with $z_\mathrm{d}=0.3\kpc$ for the thin disc and 
$z_\mathrm{d}=0.9\kpc$ for the thick disc. 
We treat the mass and the scale radius of these components, 
$(M_\mathrm{thin},  R_\mathrm{d}^\mathrm{thin}, 
  M_\mathrm{thick}, R_\mathrm{d}^\mathrm{thick})$,  
as free parameters.

\subsubsection{The atomic and molecular gas discs} 
We adopt a density profile for the atomic (H$_\mathrm{I}$) and molecular (H$_2$) gas discs of the form,
\eq{
\rho_{\operatorname{gas-disc}}(R,z) = 
\frac{\Sigma_0}{4 z_\mathrm{d}} 
\exp \left[ - \frac{R_\mathrm{m}}{R} - \frac{R}{R_\mathrm{d}} \right] 
\mathrm{sech}^2\left[ \frac{z}{2 z_\mathrm{d}} \right],
}
and fix all the parameters as in Table~1 of \cite{McMillan2017}.

\subsubsection{The dark matter halo} \label{sec:model_DM}

We adopt a flattened, generalised NFW density profile $\rho_\mathrm{DM}(R,z)$ with a suppression at large radii given by, 
\eq{
\rho_\mathrm{DM}(R,z) = 
\rho_{0} 
\frac{a^3}{m^{\gamma} (a +m)^{3-\gamma}}
\exp \left[ -\left( \frac{m}{400 \kpc} \right)^{6} \right],  
}
where $m^2 = R^2 + (z/q)^2$. 
The quantity $q$ is the flattening parameter. 
Following \cite{Bowden2016}, 
we use an auxiliary variable $U$: 
\eq{\label{eq:def_u}
U = \frac{2}{\pi}\mathrm{arctan}(q). 
}
We note that 
oblate models have $U<0.5$ ($q<1$), 
prolate models have $U>0.5$ ($q>1$), 
and 
spherical models have $U=0.5$ ($q=1$).

Throughout this paper, we only consider $q \leq 1$, 
because the current version of \agama\ \citep{Vasiliev2019} 
that is used in our paper only allows computation of orbital actions in oblate (or spherical) potentials. (We note that some studies including \citealt{Posti2019} did not take this limitation into account although they used \agama.) We note that the DM density is suppressed at large Galactocentric radii so that the total mass of the DM halo is finite. However, the way we suppress the DM density is not very important in our study, because we only use inner halo stars for our inference and therefore we can only constrain the DM density profile in the inner halo.

\subsection{Distribution function model}

\subsubsection{Action-based distribution function}

We assume that the DF of the stellar halo is given by the sum of two components, the main component and the outlier component:
\eq{
f(\vector{x},\vector{v})
= (1-\eta) f_\mathrm{main}(\vector{J}[\vector{x},\vector{v}])  + \eta f_\mathrm{outlier}(\vector{x},\vector{v}).
}
Here, $\vector{J}=(J_r, J_z, J_\phi)$ are the radial, vertical, and azimuthal action, respectively \citep{BT2008}. Each of the DFs $f$, $f_\mathrm{main}$, and $f_\mathrm{outlier}$ is separately 
normalised to unity when integrated over $(\vector{x},\vector{v})$. 
The parameter $\eta$ describes the fraction of outlier stars which is assumed to be small.

The main component has an analytic form given by,
\eq{
&f_\mathrm{main}(J_r, J_\phi, J_z) = \nonumber \\
&
\frac{C_A}{(2\pi J_0)^3} \left( \frac{h(\vector{J})}{J_0} \right)^{-\Gamma} 
\left[ 1 +  \left(\frac{g(\vector{J})}{J_0}\right) \right]^{(\Gamma - B)} 
\left[ 1+\kappa \tanh \left( \frac{J_\phi}{J_{\phi,0}} \right) \right] .
}
This double power-law model was proposed by \cite{Posti2015}
and is implemented in \agama\ \citep{Vasiliev2019}. 
We note that
\eq{
h(\vector{J}) = h_r J_r + h_z J_z + h_\phi|J_\phi|,
} and
\eq{
g(\vector{J}) = g_r J_r + g_z J_z + g_\phi|J_\phi|, 
}
are functions that govern the velocity anisotropy in the 
inner and outer part of the halo. 
The coefficients in $h(\vector{J})$ and $g(\vector{J})$ are subject to 
the following constraints: 
$0<h_i$ ($i=r, z, \phi$), $h_r + h_z + h_\phi=3$, 
$0<g_i$ ($i=r, z, \phi$), $g_r + g_z + g_\phi=3$. 
The parameter $\kappa$ determines the net rotation of the stellar halo 
and $J_{\phi,0}$ determines the scale of the angular momentum 
under which the rotation is suppressed. 
In the main analysis with the Gaia RR Lyrae stars, 
we set $\kappa=0$ and 
set $J_{\phi,0}$ to be some constant. 
This is because we observe little net rotation 
for RR Lyrae stars with known line-of-sight velocity.\footnote{
In Appendix \ref{sec:validation}, 
we fix $\kappa=0$ and 
set $J_{\phi,0}=\mathrm{constant}$ for our analysis with mock data created from smooth halo models. In contrast, we treat $(\kappa, J_{\phi,0})$ as free parameters when we analyse the mock data created from one of the cosmological hydrodynamical simulations (m12m), since the stellar halo shows a net rotation. 
}
The normalisation factor $C_A$ is defined such that the integral of $f_\mathrm{main}$ over the entire 6D phase-space is unity.  

\subsubsection{Simple distribution function for the outlier population}

Based on some tests with cosmological simulations, we found that $f_\mathrm{main}$ is flexible enough to capture the DF of the inner stellar halo. In reality, however, we expect that a small fraction of our sample stars may not be well described by $f_\mathrm{main}$. 
For example, we noticed that some objects in our RR Lyrae star catalogue have a very large  
tangential velocities, probably because they are misclassified as RR Lyrae stars (e.g., due to blending with nearby sources; 
see Section 2.2 of \citealt{Iorio2019} and references therein). Such a star would deteriorate the fit of the gravitational potential, because even a single star with extremely large velocity requires a very massive DM halo 
(since we assume that all the stars in our sample are bound to the MW).
\footnote{
For example, even if $(N-1)$ stars perfectly follow an action-based model $M$, an addition of a single unbound star would make the total likelihood of the model $M$ zero. This is because action $\vector{J}$ is not defined for an unbound star and therefore $f(\vector{x}, \vector{v})=0$ for the unbound star if $f$ is action-based. 
} In order to handle these outlier stars, we introduce $f_\mathrm{outlier}(\vector{x}, \vector{v})$ given by,
\eq{
f_\mathrm{outlier}(\vector{x}, \vector{v}) = 
\begin{cases}
0, \text{\hspace{0.81in}(if $\vector{x}$ is outside survey volume)}\\
C_B \left (2\pi \sigma_\mathrm{outlier}^2 \right)^{-3/2} 
\exp \left[  -\frac{|\vector{v}|^2}{2 \sigma_\mathrm{outlier}^2} \right], \;\text{(otherwise)}\\
\end{cases}
}
with $\sigma_\mathrm{outlier}=1000 \kms$. 
The normalisation factor $C_B$ is the reciprocal of the survey volume so that the integration of $f_\mathrm{outlier}$ over the 6D phase-space accessible to the survey is unity.

\subsection{Selection function model} \label{sec:selection_function}

We model the sample selection function in equation (\ref{eq:obs_selection})  as, 
\eq{
S (\DM, \ell, b) =
\begin{cases}
1, (\DM_\mathrm{min}(b) \leq \DM \leq \DM_\mathrm{max}, |b| \geq 20^\circ ),  \\
0, (\text{otherwise}).
\end{cases}
}
Here, $\DM_\mathrm{min}(b)$ and $\DM_\mathrm{max}$ are the minimum and maximum distance moduli for each line-of-sight, and are given by
\eq{
\DM_\mathrm{min}(b) &= 5 \log_{10} \left( \frac{5 \kpc/ |\sin b|}{10 \pc} \right), \\
\DM_\mathrm{max}    &= 5 \log_{10} \left( \frac{20 \kpc }{10 \pc} \right) = 16.505.
}
In this selection function model, it is implicitly assumed that the completeness of the RR Lyrae stars is 100\% in the survey volume. 
However, our result is unaffected as long as the completeness is constant within the survey volume. Indeed, 
according to Figure~13 of \cite{Mateu2020}, 
the completeness of the RR Lyrae sample in Gaia DR2 at $|b| > 20^\circ$ 
is almost insensitive to the $G$-magnitude at $14 < G < 18$, where most of our sample reside. 
Thus, the simple selection function model in the above equations is reasonable for our analysis.


\subsection{Error model} \label{sec:error}

Throughout this paper, we use primed variables to denote the true (error-free) quantities. 
We assume that the observational errors on $(\ell, b, \DM, \mualpha, \mudelta, \vlos)$ 
are either negligible ($\delta$ functions) or Gaussian distributed: 
\eq{
&\mathrm{Pr}(\ell | \ell', M) = \delta(\ell - \ell'), \label{eq:EllError}\\
&\mathrm{Pr}(b | b', M) = \delta(b - b'), \label{eq:BeeError}\\
&\mathrm{Pr}(\DM | \DM', M) = 
\frac{1}{\sqrt{2\pi} \sigma_{\DM}}
\exp \left[ - \frac{(\DM-\DM')^2}{2 \sigma_{\DM}^2} \right], \label{eq:DMError}\\
&\mathrm{Pr}(\vector{\mu} | \vector{\mu'}, M) = 
\frac{1}{ 2\pi |\Sigma_\mu|^{1/2}}
\exp \left[ - \frac{1}{2} (\vector{\mu}-\vector{\mu'})^\mathrm{T} \Sigma_\mu^{-1} (\vector{\mu}-\vector{\mu'})
\right], \label{eq:PMError} \\
&\mathrm{Pr}(\vlos | \vlos', M) = 
\frac{1}{\sqrt{2\pi} \sigma_\mathrm{v}}
\exp \left[ - \frac{(\vlos-\vlos')^2}{2 \sigma_\mathrm{v}^2} \right]. \label{eq:HRVError} 
}
Here, $M$ indicates our model, 
which includes the model for the observational errors. 
As mentioned in Section \ref{sec:RRLcontent},
we assume that $\sigma_{\DM}=0.240$ 
is identical for all the RR Lyrae stars. 
We fully take into account the correlated uncertainties in 
$\vector{\mu} = (\mualpha, \mudelta)$, 
and $\Sigma_\mu$ is the covariance matrix. 
We note that equation (\ref{eq:HRVError}) 
is still valid even if the $\vlos$ is not available, 
because we can set a large value of $\sigma_v$ in such a case 
(as mentioned in \citealt{McMillanBinney2013}).

\section{Likelihood of the stellar halo data} \label{sec:DFlikelihood}

As mentioned in Section \ref{sec:model}, 
the objective of our analysis is to fit the kinematic data for RR Lyrae stars with a DF model. 
Here, we derive the likelihood for the RR Lyrae data 
given a set of model parameters.

The likelihood function we adopt is similar to those
that have  already been derived and discussed in previous studies (most notably \citealt{McMillanBinney2012,McMillanBinney2013,Trick2016}). However, these previous derivations ignore, or do not properly consider, the observational errors in distances to stars. We propose a new approach to handling the distance errors by taking the advantage of the fact that all the RR Lyrae stars in our sample have approximately the same error on distance modulus.

For completeness, we start our discussion from 
the case where the stellar sample is error-free. 
Then we proceed to a more realistic case in which observational errors 
(including distance errors and missing line-of-sight velocities) are taken into account.

\subsection{Formulation with error-free data}

In the absence of the observational errors, 
given the model $M$, 
the probability that $i$th star is found 
in a Cartesian phase-space volume 
$\mathrm{d}^3\vector{x} \mathrm{d}^3\vector{v}$ 
centred at $(\vector{x}_i, \vector{v}_i)$ 
is expressed as 
\eq{
\mathrm{Pr} (\vector{x}_i, \vector{v}_i | M) 
\mathrm{d}^3\vector{x} \mathrm{d}^3\vector{v} 
&= \frac
{f( \vector{x}_i, \vector{v}_i | M) S(\vector{x}_i) \mathrm{d}^3\vector{x} \mathrm{d}^3\vector{v} }
{\int \mathrm{d}^3\vector{x} \mathrm{d}^3\vector{v} \;
f( \vector{x}, \vector{v} | M) S(\vector{x}) 
} \nonumber \\
&= \frac
{f( \vector{x}_i, \vector{v}_i | M) S(\vector{x}_i) 
\left| 
\frac
{\partial (\vector{x}, \vector{v})}
{\partial \vector{u}}
\right|_{i}
\mathrm{d}^6\vector{u} 
}
{\int \mathrm{d}^3\vector{x} \mathrm{d}^3\vector{v} \;
f( \vector{x}, \vector{v} | M) S(\vector{x}) 
} . \label{eq:6DwoError}
}
Here, 
$\vector{u}$ is the observable vector defined 
in equation (\ref{eq:observable_vector}). 
The function $S(\vector{x})$  
denotes the selection function of the survey, 
which depends on position only. 
The Jacobian is given by 
\eq{
\left| 
\frac
{\partial (\vector{x}, \vector{v})}
{\partial \vector{u}}
\right|
= \frac{\ln 10}{5} k^2 d^5 \cos b,
}
where $k=4.74047 \kms(\masyr)^{-1}$ and 
$d$ is the heliocentric distance (in $\kpc$) 
corresponding to the distance modulus $\DM$. 
The subscript $i$ in the Jacobian in equation (\ref{eq:6DwoError}) 
denotes that the quantity is evaluated 
at $(\vector{x}_i, \vector{v}_i)$.

\subsection{Formulation with observational errors}
In the presence of the observational errors, 
the expression for the probability 
$\mathrm{Pr} (\vector{x}_i, \vector{v}_i | M)$ 
becomes more complicated, 
as pointed out by \cite{Trick2016}.\footnote{
However, we note that the Jacobian factor is missing 
in equations 15-16 (in their section 2.7) 
of \cite{Trick2016}. 
}
We introduce a function $E( \vector{u} |  \vector{u}' , M )$ 
that denotes the probability that a star's 
observable vector is $\vector{u}$ 
given its true vector $\vector{u}'$ and the model $M$. 
(Remember our notation 
that a primed quantity such as $\vector{u}'$ 
denotes the true value of unprimed quantity; 
see Section \ref{sec:error}).

Given the model $M$, 
the probability that $i$th star is found 
in a observable phase-space volume 
$\mathrm{d}^6\bar{\vector{u}}$ 
centred at $\bar{\vector{u}}_i$ 
is expressed as 
\eq{
&\mathrm{Pr} (\bar{\vector{u}}_i | M) 
\mathrm{d}^6\bar{\vector{u}} = \nonumber 
\\
&\frac
{
\mathrm{d}^6 \bar{\vector{u}} \; 
S(\vector{x}( \bar{\vector{u}}_i )) 
\int \mathrm{d}^6 \vector{u}' \;
E( \bar{\vector{u}}_i |  \vector{u}' , M ) 
f( \vector{x}' ( \vector{u}' ) , \vector{v}' ( \vector{u}' ) |M) 
\left| 
\frac
{\partial (\vector{x}', \vector{v}')}
{\partial \vector{u}'}
\right|
}
{
\int \mathrm{d}^6 \vector{u} 
\int \mathrm{d}^6 \vector{u}' \; 
E( \vector{u} |  \vector{u}' , M ) 
f( \vector{x}' ( \vector{u}' ) , \vector{v}' ( \vector{u}' ) |M) 
S(\vector{x}( \vector{u} ))
\left| 
\frac
{\partial (\vector{x}', \vector{v}')}
{\partial \vector{u}'}
\right|
}.
}
For brevity, we use the simplification 
$f(\vector{u}' | M) \equiv  
f( \vector{x}' ( \vector{u}' ) , \vector{v}' ( \vector{u}' ) |M)$. 
Then, we obtain an expression for $\mathrm{Pr} (\bar{\vector{u}}_i | M)$:
\eq{\label{eq:Pr_uobs}
\mathrm{Pr} (\bar{\vector{u}}_i | M) 
=\frac
{
S(\vector{x}( \bar{\vector{u}}_i )) 
\int \mathrm{d}^6 \vector{u}' \;
E( \bar{\vector{u}}_i |  \vector{u}' , M ) 
f( \vector{u}' | M) 
\left| 
\frac
{\partial (\vector{x}', \vector{v}')}
{\partial \vector{u}'}
\right|
}
{
\int \mathrm{d}^6 \vector{u} 
\int \mathrm{d}^6 \vector{u}' \; 
E( \vector{u} |  \vector{u}' , M ) 
f( \vector{u}' | M) 
S(\vector{x}( \vector{u} ))
\left| 
\frac
{\partial (\vector{x}', \vector{v}')}
{\partial \vector{u}'}
\right|
} .
}

\subsection{Evaluation of equation (\ref{eq:Pr_uobs}) for the RR Lyrae sample}

The result of equation (\ref{eq:Pr_uobs}) 
is generally applicable to both 6D data and 5D data, 
including our RR Lyrae sample with missing $\vlos$. 
This is because 5D data without $\vlos$ data 
is equivalent to 6D data with large observational errors in $\vlos$ 
\citep{McMillanBinney2013}. 
In the following, 
we will show how to evaluate 
$\mathrm{Pr} (\bar{\vector{u}}_i | M)$ 
in equation (\ref{eq:Pr_uobs}) 
for our RR Lyrae sample.

\subsubsection{Denominator in equation (\ref{eq:Pr_uobs})} \label{sec:denominator}

By using the selection function model (Section \ref{sec:selection}) 
and the error model (Section \ref{sec:error}), 
the denominator of equation (\ref{eq:Pr_uobs}) is given by 
\eq{
&A =
\int \mathrm{d}^6 \vector{u}' 
f( \vector{u}' | M) 
\left| 
\frac
{\partial (\vector{x}, \vector{v})}
{\partial \vector{u}'}
\right|
\int \mathrm{d}^6 \vector{u}  \; 
E( \vector{u} |  \vector{u}' , M ) 
S( \DM, \ell, b )
\nonumber\\
&= 
\int \mathrm{d}^6 \vector{u}' 
f( \vector{u}' | M ) 
\left| 
\frac
{\partial (\vector{x}, \vector{v})}
{\partial \vector{u}'}
\right|
\int \mathrm{d}^6 \vector{u} \; 
S( \DM, \ell, b )
\mathrm{Pr}(\DM | \DM', M)
\nonumber\\
&\times
\delta(\ell-\ell')
\delta(b-b') 
\mathrm{Pr}(\vlos | \vlos', M)
\mathrm{Pr}(\vector{\mu}|\vector{\mu'}, M).
}
With equations (\ref{eq:EllError})-(\ref{eq:PMError}), 
the integration over 
$(\ell, b, \muell, \mub)$ 
reduces to unity. 
By using equation (\ref{eq:HRVError}) 
and assuming $\sigma_\mathrm{v} \to \infty$, 
the integration over $\vlos$ also reduces to unity. 
Thus we obtain  
\eq{ \label{eq:norm}
&A=
\int_{\text{footprint}} \mathrm{d}^2 [\ell', (\sin b')] 
\int \mathrm{d}^4 [\DM', \vector{\mu'}, \vlos'] \;
f( \vector{u}' |M) 
\nonumber\\ 
&\times 
\frac{\ln 10}{5} k^2 (d')^5 
\int_{\DM_\mathrm{min}}^{\DM_\mathrm{max}} \mathrm{d} \DM \;
\mathrm{Pr}(\DM|\DM',M)  
\nonumber\\ 
&= 
\int_{\text{footprint}} \mathrm{d}^2 [\ell', (\sin b')] 
\int \mathrm{d}^4 [\DM', \vector{\mu'}, \vlos'] \;
f( \vector{u}' |M)  
\nonumber\\ 
&\times 
\frac{\ln 10}{5} k^2 (d')^5 
\frac{1}{2}
\left[ 
\mathrm{erf} \left( \frac{\DM_\mathrm{max} - \DM'}{\sqrt{2} \sigma_{\DM}} \right) 
+
\mathrm{erf} \left( \frac{\DM' - DM_\mathrm{min}}{\sqrt{2} \sigma_{\DM}} \right) \right] .
}
Here, $(\DM_\mathrm{min}, \DM_\mathrm{max})$ 
are defined in Section \ref{sec:selection}. 
An intuitively understandable expression for $A$ can be obtained 
by performing the integration over $(\mualpha', \mudelta', \vlos')$:
\eq{\label{eq:norm_density}
&A= 
\int_{\text{footprint}} \mathrm{d}^2 [\ell', (\sin b')] 
\int \mathrm{d} \DM' \;
\rho(\DM', \ell', b' |M) 
\nonumber\\ 
&\times 
\frac{\ln 10}{5} (d')^3 
\frac{1}{2}
\left[ 
\mathrm{erf} \left( \frac{\DM_\mathrm{max} - \DM'}{\sqrt{2} \sigma_{\DM}} \right) 
+
\mathrm{erf} \left( \frac{\DM' - \DM_\mathrm{min}}{\sqrt{2} \sigma_{\DM}} \right) \right].
}
Here, 
$\rho(\DM', \ell', b' |M) = 
\int \mathrm{d}^3 \vector{v}' \; f(\vector{u}' |M)$ 
is the (normalized) stellar density.  
In the limit of $\sigma_{\DM} \to 0$, 
the factor $\frac{1}{2}[\mathrm{erf}(.)+\mathrm{erf}(.)]$ is unity 
and thus $A$ is the mass enclosed in the survey volume, 
as pointed out by \cite{Trick2016}. 
In the presence of non-zero $\sigma_{\DM}$, 
$A$ can be interpreted as the mass enclosed in the survey volume 
which is blurred by the distance errors. 
Practically, we find that the integration over $\DM'$ 
needs to be performed at 
$\DM_\mathrm{min}-4\sigma_{\DM} < \DM' < \DM_\mathrm{max}+4\sigma_{\DM}$ 
in equations (\ref{eq:norm}) and (\ref{eq:norm_density}).

In general, $\sigma_{\DM}$ is non-zero and each star has a different value of $\sigma_{\DM}$. In such a case, we need to evaluate $A$ for each star, which is computationally very expensive. This is why $\sigma_{\DM}$ is neglected (or explicitly set to be zero) in previous studies \citep{McMillanBinney2013,Trick2016}. 
However, if $\sigma_{\DM}$ is approximately the same for the entire sample, which is the case for our RR Lyrae stars, we need to evaluate $A$ only once for a given model, by assuming a single value for $\sigma_{\DM}$. 
With this prescription, 
we can dramatically reduce the computational cost 
while keeping our likelihood evaluation more precise. 
The derivation of equations (\ref{eq:norm}) and 
(\ref{eq:norm_density}) is the most important improvement 
we have over the previous formulation 
by \cite{McMillanBinney2013} and \cite{Trick2016}. 

\subsubsection{Numerator in equation (\ref{eq:Pr_uobs})}

The numerator of equation (\ref{eq:Pr_uobs}) 
can be numerically evaluated using Monte Carlo integration.

If we had 6D data for the RR Lyrae stars, 
the evaluation would be relatively easy, 
because we would only need to sample from the error distribution of $\vector{u}'$ to perform numerical integration. 
In such a case, we first randomly draw  $N_\mathrm{MC}$ realizations of the observable vector 
$\vector{u}'_{ij} =(\ell', b', \DM', \mualpha', \mudelta', \vlos')_{ij}$ 
($j=1,\cdots,N_\mathrm{MC}$) 
from the corresponding error distributions, defined in equations (\ref{eq:EllError})-(\ref{eq:HRVError}), 
centred around $\bar{\vector{u}}_i$. 
Then, by using these $N_\mathrm{MC}$ realisations of $\{\vector{u}'_{ij} \}$, 
we can evaluate the equation (\ref{eq:Pr_uobs}):
\eq{\label{eq:6D_MC_Lkhd_i}
\mathrm{(6D)} \;
\mathrm{Pr} (\bar{\vector{u}}_i | M) 
= \frac{1} {A} 
 S(\vector{x}( \bar{\vector{u}}_i )) 
\sum_{j=1}^{N_\mathrm{MC}} 
{
f(  \vector{u}'_{ij} | M ) 
\left| 
\frac
{\partial (\vector{x}', \vector{v}')}
{\partial \vector{u}'}
\right|_{ij}
} .
}
We use equation (\ref{eq:6D_MC_Lkhd_i}) to evaluate the model likelihood 
when we analyse the mock 6D data in Section \ref{sec:validation}.

For our 5D RR Lyrae sample, the above-mentioned sampling method needs a modification, because the 5D data lack in $\vlos$. For example, if we naively sample $\vlos'$ from a very wide distribution 
(assuming large $\sigma_v$ in equation (\ref{eq:HRVError})), a large fraction of the sampled phase-space coordinate $\vector{u}'_{ij}$ corresponds to unbound stars. To achieve computational efficiency, we adopt an importance sampling for $\vlos'$ using a Cauchy distribution.
Namely, for the $i$th star, we first draw $N_\mathrm{MC}$ samples from a Cauchy distribution.  
The scale parameter for the Cauchy distribution is fixed to $150 \kms$ 
and the location parameter is set to be the Solar reflex motion in the direction of the $i$th star, 
$-(\vector{v}_\odot \cdot \vector{e}_{\mathrm{los},i})$. 
The other 5D phase-space coordinates are drawn in the same manner as before, using the error distribution given in equations (\ref{eq:EllError})-(\ref{eq:PMError}). 
For each realization of the observable vector, 
$\vector{u}'_{ij} =(\ell', b', DM', \mualpha', \mudelta', \vlos')_{ij}$, 
we assign a weight 
\eq{
w_{ij} = \pi (150 \kms) 
\left[ 1 + \left( \frac{{\vlos'}_{ij} - ( - \vector{v}_\odot \cdot \vector{e}_{\mathrm{los},i} )}{150 \kms} \right)^2  \right] ,
}
which is the reciprocal of the probability density of the above-mentioned Cauchy distribution. 
Here, $\pi \simeq 3.14$ is a mathematical constant. 
Finally, by using the realisations of $\vector{u}'_{ij}$ and the weight $w_{ij}$, 
we evaluate the equation (\ref{eq:Pr_uobs}):
\eq{\label{eq:5D_MC_Lkhd_i}
\mathrm{(5D)} \;
\mathrm{Pr} (\bar{\vector{u}}_i | M) 
= \frac{1} {A} 
 S(\vector{x}( \bar{\vector{u}}_i )) 
\sum_{j=1}^{N_\mathrm{MC}} 
{
w_{ij} f( \vector{u}'_{ij} | M ) 
\left| 
\frac
{\partial (\vector{x}', \vector{v}')}
{\partial \vector{u}'}
\right|_{ij}
} .
}

\subsection{Likelihood of the RR Lyrae stars} \label{sec:totalLikelihoodRRL}

By using the expressions above, 
the logarithmic likelihood of 
the entire observed data set from the RR Lyrae sample 
given a model $M$ can be expressed as 
\eq{ \label{eq:6D_MC_Lkhd_sum_over_i}
&\text{(6D)} \sum_{i=1}^{N_\mathrm{RRL}} \ln \mathrm{Pr} (\bar{\vector{u}}_i | M) 
\nonumber \\
&=
- N_\mathrm{RRL} \ln A  
+ 
\sum_{i=1}^{N_\mathrm{RRL}} 
\ln \left[
\sum_{j=1}^{N_\mathrm{MC}} 
{
f( \vector{u}'_{ij} | M ) 
\left| 
\frac
{\partial (\vector{x}', \vector{v}')}
{\partial \vector{u}'}
\right|_{ij}
} 
\right] ,
}
in the case of 6D data and 
\eq{ \label{eq:5D_MC_Lkhd_sum_over_i}
&\text{(5D)} \sum_{i=1}^{N_\mathrm{RRL}} \ln \mathrm{Pr} (\bar{\vector{u}}_i | M) 
\nonumber \\
&=
- N_\mathrm{RRL} \ln A  
+ 
\sum_{i=1}^{N_\mathrm{RRL}} 
\ln \left[
\sum_{j=1}^{N_\mathrm{MC}} 
{
w_{ij} f( \vector{u}'_{ij} | M ) 
\left| 
\frac
{\partial (\vector{x}', \vector{v}')}
{\partial \vector{u}'}
\right|_{ij}
} 
\right] , 
}
in the case of 5D data. 
Here, we assume that 
our RR Lyrae sample stars are complete within our survey volume 
and thus 
$S(\vector{x}( \bar{\vector{u}}_i )) =1$ for all the stars 
($i=1, \cdots, N_\mathrm{RRL}$).

In our analysis, 
however, we slightly modify this likelihood and adopt a total likelihood given by 
\eq{ \label{eq:lnLRRL}
\ln L_\mathrm{RRL}( \{ \bar{\vector{u}} \} | M ) 
= \frac{N_\mathrm{RRL,eff}}{N_\mathrm{RRL}} \sum_{i=1}^{N_\mathrm{RRL}} \ln \mathrm{Pr} (\bar{\vector{u}}_i | M) ,
}
with $N_\mathrm{RRL,eff} = 1000$. 
This modification is a necessary compromise 
between the numerical accuracy requirements and limited computational resources, 
as described in the following subsection.

\subsubsection{Numerical evaluation of the total likelihood}

Our goal is 
to combine the likelihood of the RR Lyrae data given model $M$, 
$\ln L(\mathrm{Data(RRL)} | M)$, 
with the likelihood functions 
of the circular velocity data and the vertical force data 
for our MCMC analysis. 
Thus, we need to be careful  
so that the numerical noise in the likelihood 
will not seriously affect our inference of the model parameters. 
Here, 
we focus on the 5D case and 
we describe two important numerical techniques to achieve our goal.

The first technique is related to  
the Monte Carlo integration of 
equation (\ref{eq:5D_MC_Lkhd_i}). 
The precision of this integration is determined by the number of Monte Carlo samples, $N_\mathrm{MC}$. Ideally, it is desirable to set $N_\mathrm{MC}$ as large as possible 
to minimise the numerical noise in 
$\mathrm{Pr} (\bar{\vector{u}}_i | M)$.
However, this requires a large computational cost although we are not specifically interested in the {\it absolute} value of 
$\mathrm{Pr} (\bar{\vector{u}}_i | M)$. 
Rather, 
we are more interested in the {\it relative} value of 
$\mathrm{Pr} (\bar{\vector{u}}_i | M)$ 
for different models (e.g., $M=M_1$ and $M=M_2$). 
Thus, following \cite{McMillanBinney2013}, 
we use the same set of sampling points $\vector{u}'_{ij}$ and weights $w_{ij}$ 
throughout our MCMC analysis. 
We find that $N_\mathrm{MC}=100$ is enough for our purpose. 

The second technique is related to  the evaluation of $A$ in equation (\ref{eq:5D_MC_Lkhd_sum_over_i}). As discussed in Section \ref{sec:denominator}, 
the value of $A$ is common for all the RR Lyrae stars. 
If the fractional error in $A$ is $\epsilon$ ($\epsilon \ll 1$), then this error results in an error of 
$\delta (\log_{10}L(\mathrm{Data(RRL)} | M)) = (1/\ln 10) N_\mathrm{RRL,eff} \epsilon = 0.43 N_\mathrm{RRL,eff} \epsilon$ 
(see discussion in \citealt{McMillanBinney2013}). 
If we require a tolerance of $\delta (\log_{10}L) < 0.5$, 
then the fractional error in $A$ has to satisfy $\epsilon < (0.5/\ln 10) / N_\mathrm{RRL,eff} =
1/(0.87 N_\mathrm{RRL,eff})$. 
With  unlimited computational resources we could have set $N_\mathrm{RRL,eff}=16197(=N_\mathrm{RRL})$. 
However, the evaluation of $A$ (see equation (\ref{eq:norm})) is computationally challenging, 
because it involves 
6D integration in the phase-space of observable quantities 
and the conversion of observable quantities into actions. 
We use an adaptive multidimensional integration package, 
\texttt{cubature} (\url{https://github.com/stevengj/cubature}),  
to evaluate $A$, 
and find that even with this sophisticated package, 
the integration does not converge within 10 minutes per model \footnote{
If we are to run MCMC for 5000 steps (with a single MCMC `walker'), 
we need to evaluate the likelihood 5000 times.  
This would take $\sim1$ month 
assuming 10 minutes per model for evaluating likelihood. 
}
if we set $N_\mathrm{RRL,eff}=N_\mathrm{RRL}$. 
After some experiments, 
we find that setting $N_\mathrm{RRL,eff} = 1000$ 
(or $N_\mathrm{RRL,eff} \leq 3000$) 
is a reasonable choice for our analysis  
in terms of the computational speed and numerical accuracy. 
Mathematically, setting $N_\mathrm{RRL,eff} < N_\mathrm{RRL}$ 
is equivalent to assigning a weight of $(N_\mathrm{RRL,eff}/N_\mathrm{RRL})$ 
to each of our RR Lyrae stars. 
As a result, the constraining power from our RR Lyrae star sample is reduced, 
as if we only had $N_\mathrm{RRL,eff}$ stars in our catalog.


\section{Analysis}
\label{sec:analysis}

From Bayes' theorem, 
the posterior distribution of the model parameters $M$ given the data $D$ is expressed as 
\eq{
\mathrm{Pr} (M | D) = 
\frac
{\mathrm{Pr} (D| M) \mathrm{Pr} (M)}
{\mathrm{Pr} (D)} , 
}
where the Bayesian evidence, $\mathrm{Pr} (D)$, 
can be considered as a constant in our analysis. 
In this Section, 
we discuss the total likelihood $\mathrm{Pr} (D| M)$,
our choice of the prior $\mathrm{Pr} (M)$, 
and some description on the actual implementation of our Bayesian analysis.

\subsection{Bayesian likelihood}

\subsubsection{Likelihood of the circular velocity data}

The circular velocity at radius $R$ 
is given by 
\eq{
v_\mathrm{circ}^\mathrm{model} (R) = \left[ 
R \left(
\frac{\partial \Phi (R,z)}{\partial R} 
\right)
\right]^{1/2}_{z=0} .
}
By using the measured circular velocity and the associated random error 
$\{ v_\mathrm{circ}(R_{\mathrm{circ},i}) \pm \sigma_\mathrm{circ,rand}(R_{\mathrm{circ},i}) \}$  
at radius $\{ R_{\mathrm{circ},i} \}$ (for $i=1, \cdots, N_\mathrm{circ}$) 
taken from \cite{Eilers2019}, 
the logarithmic likelihood of the circular velocity data is given by 
\eq{ \label{eq:likelihoodCirc} 
\ln L_\mathrm{circ} 
= - \sum_{i=1}^{N_\mathrm{circ}} \frac{1}{2} \left( 
\frac{v_\mathrm{circ}(R_{\mathrm{circ},i}) - v_\mathrm{circ}^\mathrm{model} (R_{\mathrm{circ},i})}
{\sigma_\mathrm{circ,rand}( R_{\mathrm{circ},i} )} \right)^2 .
}

\subsubsection{Likelihood of the vertical force data}

The vertical force at $(R,z)=(R, 1.1 \kpc)$ 
is given by 
\eq{
K_{z,1.1 \kpc}^\mathrm{model} (R) = 
\left[ 
- \frac{\partial \Phi (R,z)}{\partial z} 
\right]_{z=1.1 \kpc} .
}
By using the measured vertical force and the associated error 
$\{ K_z (R_{\mathrm{Kz},i}) \pm \sigma_\mathrm{Kz}(R_{\mathrm{Kz},i}) \}$ 
at radius $\{ R_{\mathrm{Kz},i} \}$ (for $i=1, \cdots, N_\mathrm{Kz}$)
taken from \cite{BovyRix2013}, 
the logarithmic likelihood of the vertical force data is expressed as 
\eq{ \label{eq:LikelihoodKz}
\ln L_\mathrm{Kz}  
= - \sum_{i=1}^{N_\mathrm{Kz}} \frac{1}{2} \left( 
\frac{K_z( R_{\mathrm{Kz},i} ) - K_z^\mathrm{model} ( R_{\mathrm{Kz},i} )}
{\sigma_\mathrm{Kz}( R_{\mathrm{Kz},i} )} \right)^2 . 
}
%

\newcommand{\tableTwoCaption}{
Bayesian prior and posterior distributions of our model parameters and the best-fitting parameters in our main analysis with RR Lyrae stars. 
}
\newcommand{\tableTwoNote}{
Note: 
(1) $U=(2/\pi)\mathrm{arctan}(q)$ [equation (\ref{eq:def_u})]. 
The range of $0.2 < U \leq 0.5$ corresponds to $0.3249 < q \leq 1$. 
(2) These quantities are used as priors. See Section \ref{sec:prior}.
(3) See equation (\ref{eq:hgprior}).
(4) $(\kappa, J_{\phi,0})$ are fixed in this paper, 
except for the mock analysis with mock data generated from a cosmological simulation (see Appendix \ref{sec:m12m_mock}). 
}
\begin{table*}
\centering
\caption{ {\tableTwoCaption} }
\label{tab:maxL}
\begin{tabular}{llllll} 
\hline
Fit parameter & Quantity & Prior distribution & Note & Posterior distribution & Best-fitting\\
\hline
\hline 
Free &
$M_\mathrm{bulge}$ & $M_\mathrm{bulge} = (8.9\pm0.89) \times 10^9 M_\odot$ & ... & 
$[8.33, 9.23, 10.09]\times 10^9 M_\odot$ & $9.53 \times 10^9 M_\odot$ \\

Free &
$M_\mathrm{thin}$ & $M_\mathrm{thin} = (35 \pm 10) \times 10^9 M_\odot$ & ... & 
$[34.39, 36.53, 38.55]\times 10^9 M_\odot$ & $37.22\times 10^9 M_\odot$\\

Free &
$M_\mathrm{thick}$ & $M_\mathrm{thick} = (6 \pm 3) \times 10^9 M_\odot$ & ... & 
$[4.45, 6.21, 8.59]\times 10^9 M_\odot$ & $7.30\times 10^9 M_\odot$\\

Free &
$R_\mathrm{d}^\mathrm{thin}$  & $R_\mathrm{d}^\mathrm{thin}= (2.6\pm0.5) \kpc$  & ... & 
$[2.58, 2.68, 2.80] \kpc$ & $2.63 \kpc$\\

Free &
$R_\mathrm{d}^\mathrm{thick}$ & $R_\mathrm{d}^\mathrm{thick}= (2.0\pm0.2) \kpc$  & ... & 
$[1.76, 1.96, 2.15] \kpc$ & $1.94 \kpc$\\

Fixed &
$z_\mathrm{d}^\mathrm{thin}$  & $z_\mathrm{d}^\mathrm{thin}= 0.3 \kpc$ (fixed)  & ... & 
$0.3 \kpc$ (fixed) &... \\

Fixed &
$z_\mathrm{d}^\mathrm{thick}$ & $z_\mathrm{d}^\mathrm{thick}= 0.9 \kpc$ (fixed)  & ... & 
$0.9 \kpc$ (fixed) &...  \\

Free &
$q$ & Flat at $0.2 < U \leq 0.5$ (oblate) & (1) & 
$[0.983, 0.993, 0.998]$ & $0.996$\\

Free &
$\gamma$ & Flat at $0 < \gamma < 1.9$ & ... &
$[0.785, 0.982, 1.209]$ & $0.738$\\

Free &
$a$ & Flat at $-\infty < \log_{10} a < \infty$ & ... &
$[10.29, 12.49, 16.66] \kpc$ & $10.46 \kpc$\\

Free &
$\log_{10} \rho_0$ & Flat at $-\infty < \log_{10} \rho_0 <\infty$ & ... &
$[6.89, 7.21, 7.44]$ & $7.44$ \\

... &
Thick to thin disk ratio  & $\Sigma_\mathrm{thick} (R_0) / \Sigma_\mathrm{thin} (R_0)  =(0.12\pm0.04)$ & (2) & [0.070, 0.104, 0.139] & $0.120$\\

... &
Total stellar mass & Equation (\ref{eq:stellarMass}) & (2) & $[50.13, 52.32, 54.24]\times 10^9 M_\odot$ & $54.04\times 10^9 M_\odot$\\

... &
Dark matter concentration $c'$ & $\ln c'_\mathrm{v} = \ln (r_{94}/r_{-2}) = 2.56 \pm 0.272$ & (2) & $[18.45, 19.55, 20.89]$ & $18.87$\\
\hline 

Free &
$a_\mathrm{in}$ & Flat at $0<a_\mathrm{in}<1$. & (3) & [0.234, 0.404, 0.649] & $0.331$\\

Free &
$b_\mathrm{in}$ & Flat at $0<b_\mathrm{in}<1$. & (3) & [0.789, 0.879, 0.921] & $0.890$\\

Free &
$a_\mathrm{out}$ & Flat at $0<a_\mathrm{out}<1$. & (3) & [0.292, 0.342, 0.383] & $0.346$\\

Free &
$b_\mathrm{out}$ & Flat at $0<b_\mathrm{out}<1$. & (3) & [0.792, 0.836, 0.861] & $0.842$\\

Free &
$\log_{10} J_0$ & Flat at $-\infty  < \log_{10} J_0 < \infty $ & ... & [3.02, 3.22, 3.59] & $3.25$\\

Free &
$\Gamma$  & Flat at $0 \leq \Gamma < 2.8$ & ... & [0.49, 1.47, 2.47] & $1.50$\\

Fixed &
$B$       & Fixed to $B=5$ & ... & $5$ (fixed) & ...  \\

Fixed &
$\kappa$       & Fixed to $\kappa=0$ (non-rotating) & (4) & $0$ (fixed) & ...  \\

Fixed &
$J_{\phi,0}$       & Fixed to $J_{\phi,0}=\mathrm{const}.$ & (4) & $\mathrm{const}.$ (fixed) & ...  \\

Free &
$\log_{10} \eta$ & Flat at $-\infty < \log_{10} \eta < -2$ & ... & $[-7.26, -4.65, -3.21]$ & $-3.99$ \\
\hline 
\end{tabular} \\
\flushleft{\tableTwoNote}
\end{table*}


\subsubsection{Total likelihood of the data}

Given the model $M$, 
the logarithmic likelihood of the entire data $D$ is expressed as 
\eq{
\ln \mathrm{Pr}(D|M) 
&= \ln L_\mathrm{circ} + \ln L_\mathrm{Kz} + \ln L_\mathrm{RRL} , 
}
where $L_\mathrm{circ}$, $L_\mathrm{Kz}$, and $L_\mathrm{RRL}$ 
are defined in equations 
(\ref{eq:likelihoodCirc}), (\ref{eq:LikelihoodKz}), and (\ref{eq:lnLRRL}), 
respectively. 
We note that 
we reduce the weight from RR Lyrae data by a factor 
$(N_\mathrm{RRL,eff}/N_\mathrm{RRL}) = 1000/16197$ 
when we compute $L_\mathrm{RRL}$ due to our computational limitation 
(see Section \ref{sec:totalLikelihoodRRL}).

\subsection{Bayesian prior} \label{sec:prior}

Our Bayesian prior is summarised in Table \ref{tab:maxL}. 
We note that the prior distributions 
for the parameters of the model potential 
are mostly taken from \cite{BlandHawthorn_Gerhard2016} 
and \cite{McMillan2017}. 
We have tried various prior distributions  
and confirmed that 
the choice of the prior distribution 
does not change the main conclusion of our paper, 
especially the flattening of the DM halo.

Because we do not allow prolate DM distributions, 
we use a flat prior for $U$ (see equation (\ref{eq:def_u})): $0.2 < U \leq 0.5$. 
(This range of $U$ corresponds to $0.3249 < q \leq 1$.) 
We use similar definitions for 
the priors on the concentration parameter, virial mass, and virial radius 
to those in \cite{McMillan2017}, 
by evaluating these quantities by spherically averaging the non-spherical density.
Namely, for a given set of parameters 
$(\rho_0, \gamma, a, q)$, 
we define two radii $r_{200}$ and $r_{94}$ 
such that the mean density within a sphere of $r_{200}$ and $r_{94}$ 
is 200 and 94 times the critical density 
($\rho_\mathrm{crit} = 137.55 \; M_\odot \kpc^{-3})$; 
see \citealt{BT2008}), 
respectively. 
The virial mass $M_{200}$ and $M_{94}$ are 
defined as the enclosed mass within the sphere of radius $r_{200}$ and $r_{94}$, respectively. 
We define $\langle \rho \rangle (r)$ 
as the mean density within a spherical shell centred at a radius $r$;  
and we define the radius $r_{-2}$ such that 
$\mathrm{d} \ln \langle \rho \rangle / \mathrm{d} \ln r = -2$. 
The concentration parameter is defined as 
$c'_\mathrm{v} = r_{94}/r_{-2}$.

Following \cite{McMillan2017}, we set a prior on the total stellar mass 
$M_\mathrm{star}$ (the sum of the bulge, thin disc, and thick disc)
\eq{\label{eq:stellarMass}
\log_{10}M_\mathrm{star} = 
\log_{10} \left\{ 
\frac{ M_{200} \times 2 \times 0.0351} { 
 \left(\frac{M_{200}}{M_\mathrm{knee}}\right)^{-1.376} 
+\left(\frac{M_{200}}{M_\mathrm{knee}}\right)^{ 0.608} 
}
\right\} \pm 0.20 , 
}
with $M_\mathrm{knee} = 10^{11.59}M_\odot$.

Also, we set priors on the ratio of surface densities of thin and thick discs evaluated at Solar cylinder:
\eq{
\Sigma_\mathrm{thick} (R_0) / \Sigma_\mathrm{thin} (R_0)  =(0.12\pm0.04),
}
which is taken from \cite{BlandHawthorn_Gerhard2016}. 
We note that \cite{McMillan2017} put a prior on the ratio of densities (rather than surface densities) of thin and thick discs, 
$
\rho_{\operatorname{stellar-disc}}^\mathrm{thick}(R_0, 0) / 
\rho_{\operatorname{stellar-disc}}^\mathrm{thin} (R_0, 0) =0.12\pm0.012. 
$
However, we choose the surface density ratio, 
because this quantity is more robustly estimated in the literature \citep{BlandHawthorn_Gerhard2016}.

The positive parameters $(h_r, h_z, h_\phi)$ and $(g_r, g_z, g_\phi)$ 
that govern the stellar halo velocity anisotropy  
satisfy relationships $h_r + h_z + h_\phi = 3$ and $g_r + g_z + g_\phi = 3$. 
To sample these parameters in an unbiased manner,
we draw four independent random numbers
$(a_\mathrm{in},b_\mathrm{in},a_\mathrm{out},b_\mathrm{out})$ 
from a uniform distribution between $0$ and $1$, and assign 
\eq{ \label{eq:hgprior}
\begin{cases}
h_r    &= 3 \sqrt{a_\mathrm{in}} (1-b_\mathrm{in}) \\
h_z    &= 3 \sqrt{a_\mathrm{in}} b_\mathrm{in} \\
h_\phi &= 3 (1 - \sqrt{a_\mathrm{in}}) \\
\end{cases} 
\;\;
\begin{cases}
g_r    &= 3 \sqrt{a_\mathrm{out}} (1-b_\mathrm{out}) \\
g_z    &= 3 \sqrt{a_\mathrm{out}} b_\mathrm{out} \\
g_\phi &= 3 (1 - \sqrt{a_\mathrm{out}}) .
\end{cases}
}
This method is mathematically equivalent to 
sampling points uniformly from 
a 2D region enclosed by an equilateral triangle. 

The parameter $B$ that governs the density slope in the outer parts of the stellar halo 
is fixed to $B=5$ in our analysis, 
because it is difficult to constrain $B$ with our inner halo data at $r \lesssim 30 \kpc$.  
The fraction of the outlier population $\eta$ 
is expected to be small, 
and thus we adopt a flat prior at $\log_{10} \eta<-2$.

\subsection{Markov Chain Monte Carlo analysis}

To estimate the model parameters, 
we first 
search for the maximum-likelihood parameters with a Nelder-Mead optimisation package \texttt{constrNMPy} 
(\url{https://github.com/alexblaessle/constrNMPy}) with some reasonable tolerance level. Then we use the resultant parameters as the initial condition of the Bayesian Markov Chain Monte Carlo (MCMC) analysis. We use a package \emcee\ \citep{emcee} for the MCMC analysis. We use $(2\times N_\mathrm{free})$ walkers 
(where $N_\mathrm{free}$ is the number of free parameters)  
and run the MCMC for several thousand steps. We discard initial half of the chain for burn-in 
and analysed the remaining chain.

The analysis code is written in Python, and it is developed from an example code in \agama\  
(\url{https://github.com/GalacticDynamics-Oxford/Agama/blob/master/py/example_df_fit.py}). 
In Appendix \ref{sec:validation}, we validate our method with mock data sets.

\begin{figure}
\centering
\includegraphics[width=3.2in]{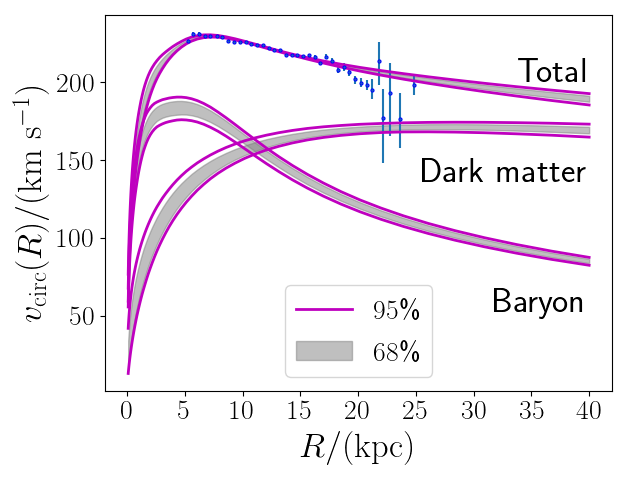}
\caption{
The radial profile of the circular velocity $v_\mathrm{circ}(R)$, 
along with its contribution from baryon and dark matter. 
The blue data points with error bar are taken from 
\protect\cite{Eilers2019}. 
The grey shaded region corresponds to the central 68 percentile 
of the posterior distribution of our model, 
while the magenta solid lines cover the central 95 percentile. 
}
\label{fig:circ}
\end{figure}

\begin{figure}
\centering
\includegraphics[width=3.2in]{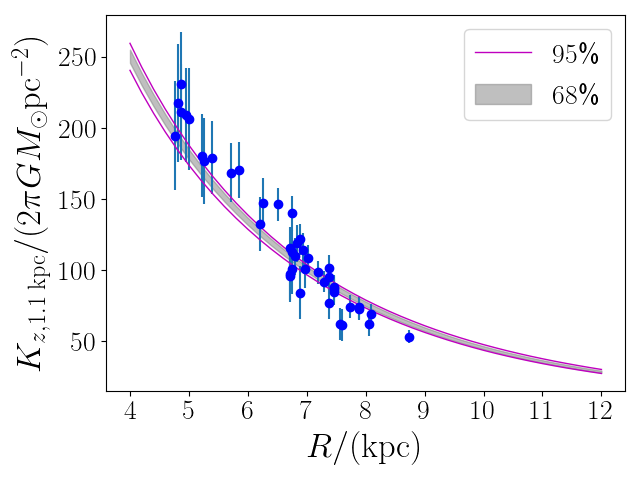}
\caption{
The radial profile of the vertical force $K_{z, 1.1 \kpc}$ 
measured at $z=1.1 \kpc$. 
The blue data points with error bar are taken from 
\protect\cite{BovyRix2013}. 
The grey shaded region corresponds to the central 68 percentile 
of the posterior distribution of our model, 
while the magenta solid lines cover the central 95 percentile.
}
\label{fig:Kz}
\end{figure}

\begin{figure*}
\centering
\includegraphics[trim= 10 0 10 60, clip, width=0.495\textwidth]{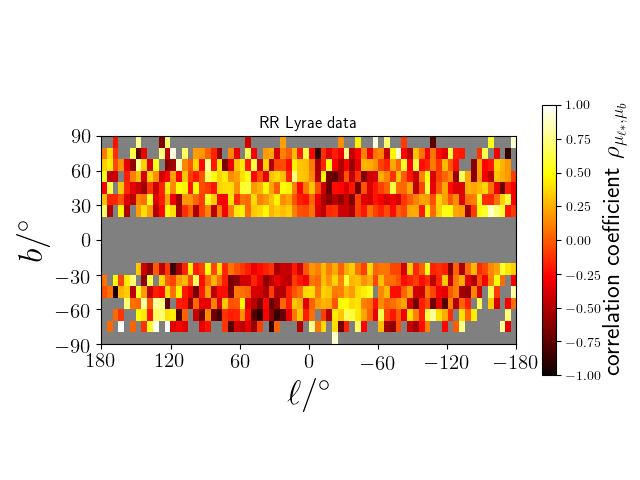}
\includegraphics[trim= 10 0 10 60, clip, width=0.495\textwidth]{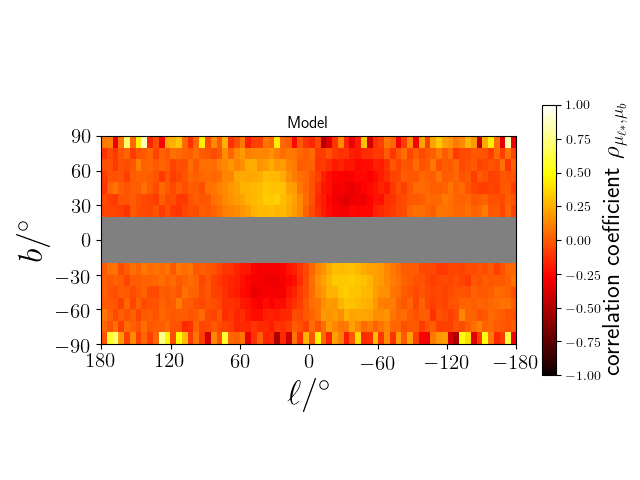}\\
\vspace{-1cm}
\includegraphics[trim= 10 0 10 60, clip, width=0.495\textwidth]{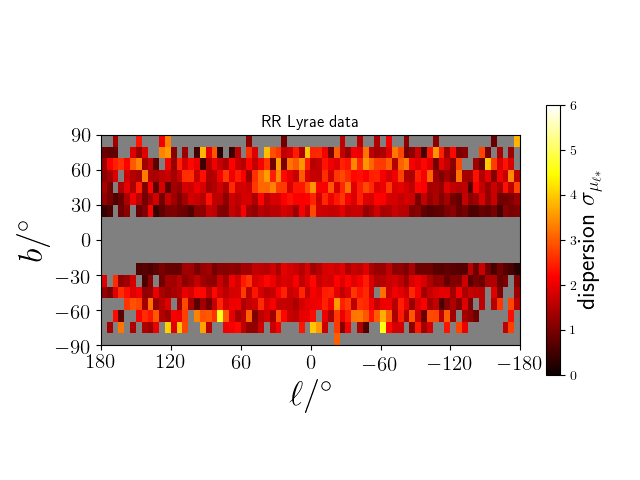}
\includegraphics[trim= 10 0 10 60, clip, width=0.495\textwidth]{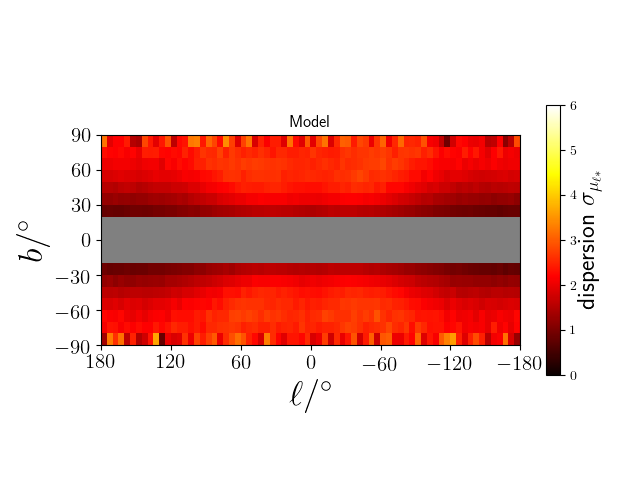}\\
\vspace{-1cm}
\includegraphics[trim= 10 0 10 60, clip, width=0.495\textwidth]{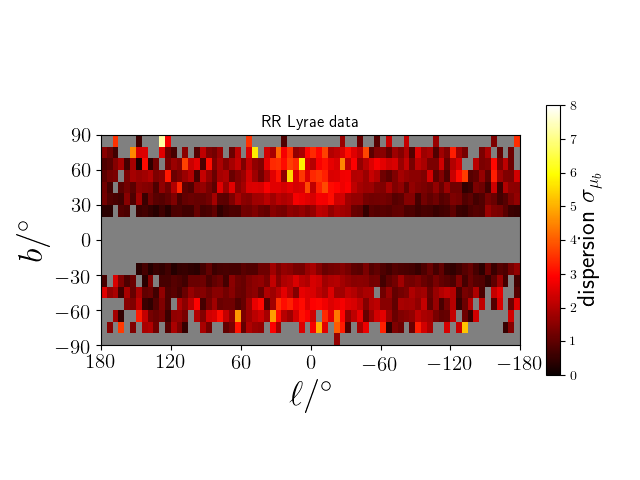}
\includegraphics[trim= 10 0 10 60, clip, width=0.495\textwidth]{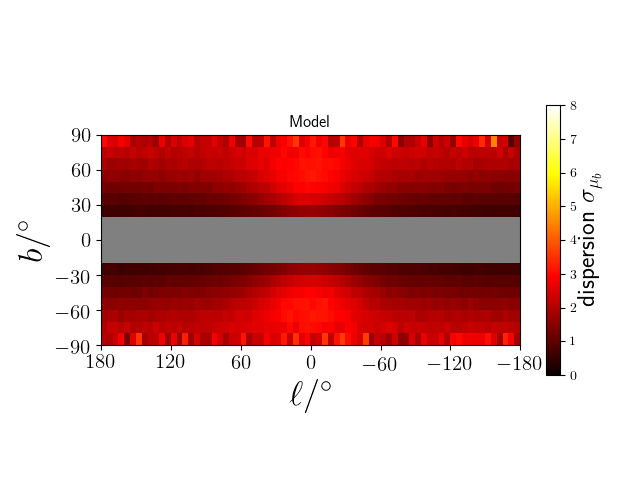}\\
\vspace{-1cm}
\caption{
A visualisation of the proper motion distribution 
of our RR Lyrae sample ({\bf left column}) 
and the average prediction of our models 
constructed from the MCMC chain ({\bf right column}). 
In the top row, 
we show the Pearson correlation coefficient 
$\rho_{\mu_{\ell *}, \mu_b}$ 
for the distribution of 2D proper motion $(\mu_{\ell *}, \mu_b)$. 
In the middle row, 
we show the dispersion $\sigma_{\mu_{\ell *}}$ 
of the distribution of $\mu_{\ell,*}$. 
In the bottom row, 
we show the dispersion $\sigma_{\mu_{b}}$ 
of the distribution of $\mu_{b}$. 
}
\label{fig:pmlb}
\end{figure*}

\section{Results} \label{sec:result}

In this Section we describe the results of our Bayesian MCMC analysis. The posterior distribution is summarised in Table \ref{tab:maxL}. 
The corner plots are given in Appendix \ref{appendix:corner}. 

\subsection{Comparison of the input data and our model}

To check the performance of our method, we first compare the input data and our model predictions. 

\subsubsection{Circular velocity} 

Fig.~\ref{fig:circ} shows the 
radial profile of the circular velocity $v_\mathrm{circ}(R)$
and the contribution from baryonic and DM components 
sampled from our posterior distribution. 
We can see that our model properly fits 
the rotation curve data from \cite{Eilers2019}. 
Also, our result suggests that the contribution from the DM surpasses that from baryon beyond $R \simeq 11 \kpc$. This transition radius is located further outward than that in \cite{deSalas2019}, which uses the same circular velocity data (but not the other datasets used here).

\subsubsection{Vertical force} 

Fig.~\ref{fig:Kz} shows the radial profile of the vertical force $K_{z,1.1\kpc}$ 
sampled from our posterior distribution. 
We can see that our model properly fits 
the data points of \cite{BovyRix2013}. 
Our posterior distribution suggests that 
the local value of $K_{z,1.1\kpc}$ 
measured at $R=R_0=8.178 \kpc$ 
is $K_{z,1.1\kpc}(R_0) = (72.7 \pm 1.4) / (2\pi G M_\odot \pc^{-2})$, 
which is consistent with the classical measurement of $(71 \pm 6) / (2\pi G M_\odot \kpc^{-2})$ by \cite{KuijkenGilmore1991}.

\subsubsection{Proper motion distribution} 

In the left hand column of Fig.~\ref{fig:pmlb}, we show the statistical properties of the proper motion distribution, 
$(\rho_{\muell, \mub}, \sigma_{\muell}, \sigma_{\mub})$, 
as a function of $(\ell, b)$ for our RR Lyrae sample. 
In computing these quantities, 
we first divide the RR Lyrae star sample into $72\times18$ cells 
with a size of $(\Delta\ell, \Delta b) = (5^\circ, 10^\circ)$. 
Then we analyse the proper motion distribution in each cell to evaluate 
$(\rho_{\muell, \mub}, \sigma_{\muell}, \sigma_{\mub})$.
\footnote{
We note that these quantities are derived from the covariance matrix of the proper motion distribution along each line-of-sight, which is different from the covariance matrix of the observational error for each star $\Sigma_\mu$ 
used in equation (\ref{eq:PMError}). 
}

In the right hand column of Fig.~\ref{fig:pmlb}, 
we also derive the same statistical properties  
by using the posterior distribution. 
First, we randomly select 160 models from our MCMC chain. 
For each model, we generate a large enough sample of mock stars from the DF, 
and add the observational uncertainty. 
Then we randomly select 16197 error-added mock stars 
within the survey volume as in our RR Lyrae sample. 
We compute 
$(\rho_{\muell, \mub}, \sigma_{\muell}, \sigma_{\mub})$ 
for each model,
and average these quantities over 160 models along each line-of-sight.

We see that the 
overall trend in the proper motion distribution is well recovered by our models. 
At  $-90^\circ$ $< \ell <$ $90^\circ$ 
and $-90^\circ$ $< b <$ $90^\circ$, 
we can see the quadrupole pattern in the correlation coefficient $\rho_{\muell, \mub}$  in the top row of Fig.~~\ref{fig:pmlb}
for both the RR Lyrae sample and our models. 
This pattern is known to be a characteristic of a radially biased velocity distribution \citep{Iorio2019}. 
However, we note that it the first time this proper motion distribution is successfully fit and recovered by a DF model. 
The model distributions for the dispersions $\sigma_{\mu_l*}$ and $\sigma_{\mu_b}$ in the next two rows also show very good overall agreement.

\subsection{Comparison of the external data and our model}

In this Section we compare our results with other independent data sets that are not used to constrain our model. This comparison serves to test the predictive power of our model.

\subsubsection{Velocity dispersion} 

In Fig.~\ref{fig:veldisp}, we show 
the radial profile of the 3D velocity dispersion 
$(\sigma_r, \sigma_\theta, \sigma_\phi)$ 
and the velocity anisotropy 
$\beta = 1 - (\sigma_\theta^2+\sigma_\phi^2)/(2\sigma_r^2)$ 
for halo K giants obtained by the LAMOST survey \citep{Bird2019}. 
These data (shown as coloured open symbols) are not used in our analysis, but are compared with our model predictions (solid and dashed lines).

For this analysis, we first randomly select 160 models from our MCMC chain. 
For each model, we generate a large enough sample of mock stars from the DF, 
without adding any observational error. 
We select those error-free 6D mock data within our survey volume 
and compute $(\sigma_r, \sigma_\theta, \sigma_\phi, \beta)$ 
for each model. 
In Fig.~\ref{fig:veldisp}, we plot the radial profile of these quantities.

The radial profiles of 
$(\sigma_r, \sigma_\theta, \sigma_\phi, \beta)$ 
for our models and 
those of K giants are broadly consistent with each other. 
In particular, 
both our models and the K giants data suggest 
highly radially-biased velocity distribution 
with $\beta \gtrsim 0.75$ at $10 \lesssim r/\kpc \lesssim 22$. 
(We note that \citealt{Bird2019} 
mentioned that $\beta(r)$ of K giants 
shows a mild drop at $r \gtrsim 25 \kpc$ 
due to the presence of substructure.) 
The high value of $\beta$ is consistent with 
the result in \cite{Belokurov2018}, 
in which work they proposed that the inner halo is 
dominated by the stellar debris of a radial merger with a massive satellite (now referred to as ``Gaia-Enceladus'' or ``Gaia-Sausage'') about $8-10$~Gyr ago 
(see also \citealt{Helmi2018Nature}). 
We note that our estimate of 
$\sigma_r$ is systematically offset from 
the observed trend of K giants at $15 \kpc < r$, 
which might reflect different survey volumes 
of the LAMOST K giants and our RR Lyrae star sample.

Our estimate of the velocity anisotropy $0.7\lesssim \beta \lesssim 0.9$ is 
significantly larger than the reported value of 
$\beta \lesssim 0.3$ for K giants in SDSS catalogue \citep{Das2016a} 
or blue horizontal branch (BHB) stars in SDSS catalogue \citep{Das2016b}. 
One possible explanation for this discrepancy is that 
they took into account the metallicity dependence of the DF while we do not. 
However, this does not fully explain the discrepancy. 
It has been known that the metal-rich part of the stellar halo shows higher value of $\beta$ (e.g., \citealt{Deason2011, Hattori2013, Kafle2013}), 
but even the metal-rich part of the DF in \cite{Das2016a} suggest $\beta \simeq 0.3$ in the inner halo, which is much smaller than our estimate of $\beta$ for the entire RR Lyrae population. 
Another possible explanation is that the proper motion data they used in \cite{Das2016a} and \cite{Das2016b} may not be accurate enough to estimate $\beta$. For example, if the proper motion errors in SDSS were underestimated, then the velocity ellipsoid could have been sphericalised  (due to insufficient deconvolution of the proper motion error), which could result in smaller value of $\beta$. 
Yet another possibility is a counter-intuitive scenario that the velocity distribution depends on the stellar type. Although we do not aggressively advocate this possibility, it may be worth noting that \cite{Utkin2020} recently claimed that the value of $\beta$ for BHB stars is typically smaller than that of RR Lyrae stars.

\subsection{Dark matter distribution}

We now discuss the properties of the halo DM distribution within $r \lesssim 30 \kpc$ as inferred from our analysis. Table~ \ref{tab:DM_summary} 
summarises the characteristic parameters of the DM density profile derived from our analysis. The correlations between some of these quantities are shown 
in Figures \ref{fig:corner_raw_all} and \ref{fig:corner_alt_all}.

\subsubsection{Dark matter density flattening} 

Fig.~\ref{fig:GaiaDMdensity}(a) shows the posterior distribution of the DM density flattening $q$. We can see that the posterior distribution is strongly peaked near $q=1$. Since $q=1$ is the upper boundary of the prior distribution, we cannot rule out the possibility 
that the DM density is prolate. The fact that 
99\% of the posterior distribution of $q$ is located above $q=0.963$ strongly disfavours even a moderately flattened DM halo. 
It is worth noting that the shape of the posterior distribution of $q$ shown in Fig. \ref{fig:GaiaDMdensity}(a) is naturally expected if the shape of the DM halo is actually nearly spherical. For example, in Appendix \ref{sec:smooth_mock}, we see that the posterior distributions of $q$ derived from our mock analysis look very similar to Fig. \ref{fig:GaiaDMdensity}(a), when the mock data is generated from a MW model with $q=0.996$ 
(see Fig. \ref{fig:DMdensity_validation_q}(c)(f)).

Recently, \cite{Wegg2019} estimated the shape of the DM density profile by applying the axisymmetric from of the Jeans equations for the kinematic data for 15651 RR Lyrae stars at $r<20 \kpc$. They also find that the shape of the DM halo is nearly spherical, with the density flattening of $q=1.00 \pm 0.09$, which is consistent with our result.

\subsubsection{Dark matter density profile} 

Fig.~\ref{fig:GaiaDMdensity} (b) and (c) 
show the DM density profile evaluated at $(R,z)=(R, 0 \kpc)$ and at $(R,z)=(R_0, z)$, respectively. (We note $R_0 = 8.178 \kpc$ is assumed.) These profiles are sampled from the posterior distribution. We can see that our model puts a tight constraint on the radial and vertical density profiles. However, we need to be careful in interpreting this result. 
First, 99\% of the posterior distribution is 
distributed at $0.963 \leq q \leq 1$, 
so the DM density profiles sampled from our posterior distribution are very close to spherical. Secondly, we use halo tracers that are distributed at $5 \lesssim r/\kpc \lesssim 30$. Thus, the inference of the DM density outside this range is less reliable. Third, 
the seemingly small variation in the density profile at large $R$ and large $|z|$ are probably because we fix the outer density slope of the DM to be $\simeq (-3)$ in the outer halo.

In spite of the above-mentioned complexities, 
we think our estimate of $\rho_\mathrm{DM}(R,0)$ is reliable for $1 \lesssim R/\kpc < 30$ based on our mock analysis in Appendix \ref{sec:validation}. This can be understood in the following manner. 
Typical halo stars have a radially elongated orbits ($\beta \simeq 0.8$; see Fig. \ref{fig:veldisp}), and many halo stars have a relatively small pericentric radii. Therefore, many halo stars in our sample have orbits that are affected by the DM distribution in the inner few kpc and therefore their kinematics, even at large radii, convey this information. 

\subsubsection{Dark matter density slope} 

In our analysis, DM's inner density slope $(-\gamma)$ is a free parameter, while the outer density slope is fixed to be $(-3)$. 
The central 68\% of the posterior distribution of $\gamma$ is distributed at $0.785 < \gamma < 1.209$ (see Table \ref{tab:DM_summary}), 
which is consistent with a cusped NFW profile with $\gamma=1$ \citep{NFW1997}. 
Fig.~\ref{fig:GaiaDMdensity}(d) 
shows the DM logarithmic density slope 
$\mathrm{d} \ln \rho_\mathrm{DM}/ \mathrm{d} \ln r$ 
as a function of Galactocentric radius $r$ evaluated at $(R,z)=(r,0)$. 
As we can see from this figure, 
the logarithmic density slope is as sharp as $(-1)$ 
at $r \simeq 1 \kpc$.

\subsubsection{Other constraints on the dark matter distribution} 

Table \ref{tab:DM_summary} summarises the key properties of the DM density profile 
derived from our analysis. 
For example, our result constraints the local DM density to, 
$\rho_{\mathrm{DM},\odot} = (9.01^{+0.18}_{-0.20})\times10^{-3}M_\odot \pc^{-3}$,
which is equivalent to 
$0.342^{+0.007}_{-0.007}$ 
$\;\mathrm{GeV} \mathrm{cm}^{-3}$. 
This result is consistent with previous measurements 
\citep{Read2014, BlandHawthorn_Gerhard2016} including the most recent global measurements based on Gaia DR2 rotation curve data \citep{deSalas2019, Cautun2020}.

We also derive the enclosed mass of the DM and the enclosed mass of the DM plus baryon at various radius $r$. For example, 
our result indicates that the DM mass within the virial radius $r_{200}$ 
is 
$M_{200} = 0.730^{+0.046}_{-0.052} \times 10^{12} M_\odot$, 
which is consistent with some other literature including \cite{McMillan2017}, but slightly lower than $(9.0 \pm 1.3)\times 10^{11}$ recently found by \cite{Vasiliev2020Tango}. 
However, we note that this result may be dominated by our prior on $M_\mathrm{star}/M_{200}$, since our sample is distributed in the inner part of the halo at $5 \lesssim r/\kpc \lesssim 30$.

\newcommand{\tableFourCaption}{
Summary of the dark matter properties of the Milky Way. 
}
\newcommand{\tableFourNote}{
}
\begin{table}
\centering
\caption{ {\tableFourCaption} }
\label{tab:DM_summary}
\begin{tabular}{ll} 
\hline
Quantities & 
[16, 50, 84] percentiles \\
\hline
\hline 
$\rho_{\mathrm{DM},\odot}$ [$M_\odot \pc^{-3}$] & 
$[0.00881, 0.00901, 0.00919]$ $M_\odot \pc^{-3}$ \\
$\rho_{\mathrm{DM},\odot}$ [GeV $\mathrm{cm}^{-3}$] & 
$[0.335, 0.342, 0.349]$ GeV $\mathrm{cm}^{-3}$ \\
$M_{200} $ &
$[0.678, 0.730, 0.776] \times 10^{12}M_\odot$ \\
$M_{94} $ &
$[0.774, 0.837, 0.894] \times 10^{12}M_\odot$ \\
$r_{200} $ &
$[180.52, 185.03, 188.84] \kpc$ \\
$r_{94}$ &
$[242.71, 249.08, 254.64] \kpc$ \\
$c' = r_{94} / r_{-2}$ &
$[18.45, 19.55, 20.89]$  \\
$r_{-2}$ &
$[11.69, 12.72, 13.73] \kpc$ \\
$a$ &
$[10.29, 12.49, 16.66] \kpc$ \\
$\gamma$ &
$[0.785,0.982, 1.209]$ \\
$q$ &
$[0.983, 0.993, 0.998]$ \\

\hline 
$M_\mathrm{DM}(r<20 \kpc)$ & 
$[0.132, 0.134, 0.137] \times 10^{12}M_\odot$ \\
$M_\mathrm{DM}(r<50 \kpc)$ &
$[0.311, 0.322, 0.330] \times 10^{12}M_\odot$ \\
$M_\mathrm{DM}(r<100 \kpc)$ & 
$[0.497, 0.523, 0.543] \times 10^{12}M_\odot$ \\
$M_\mathrm{DM}(r<200 \kpc)$ & 
$[0.711, 0.759, 0.798] \times 10^{12}M_\odot$ \\
$M_\mathrm{DM}(r<300 \kpc)$ &
$[0.845, 0.907, 0.960] \times 10^{12}M_\odot$ \\
\hline 
$M_\mathrm{total}(r<20 \kpc)$ & 
$[0.182, 0.186, 0.191] \times 10^{12}M_\odot$ \\
$M_\mathrm{total}(r<50 \kpc)$ &
$[0.361, 0.374, 0.384] \times 10^{12}M_\odot$ \\
$M_\mathrm{total}(r<100 \kpc)$ & 
$[0.547, 0.575, 0.598] \times 10^{12}M_\odot$ \\
$M_\mathrm{total}(r<200 \kpc)$ & 
$[0.761, 0.811, 0.852] \times 10^{12}M_\odot$ \\
$M_\mathrm{total}(r<300 \kpc)$ &
$[0.895, 0.959, 1.015] \times 10^{12}M_\odot$ \\
\hline 
\end{tabular} \\
\flushleft{\tableFourNote}
\end{table}


\begin{figure}
\centering
\includegraphics[width=3.2in]{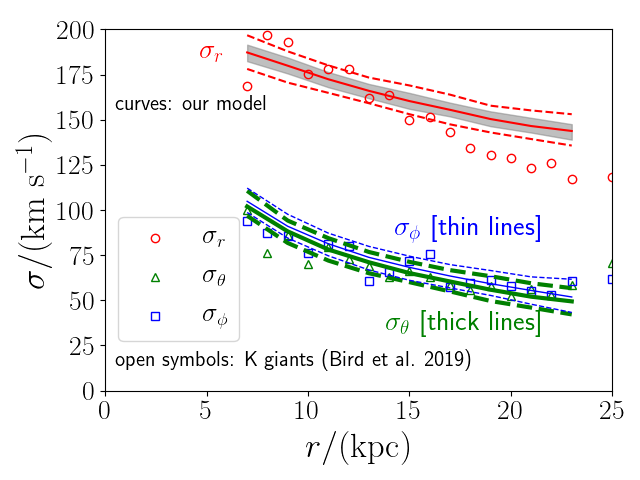} \\
\includegraphics[width=3.2in]{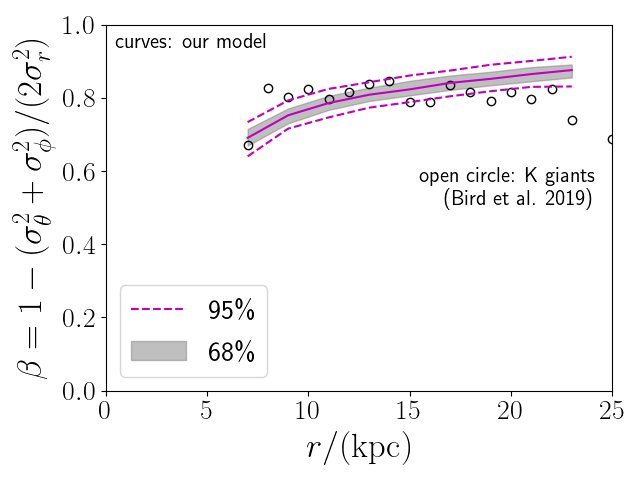}
\caption{
({\bf Top}): 
The radial profile of the velocity dispersion 
$\sigma_r$ (red),
$\sigma_\theta$ (green), and 
$\sigma_\phi$ (blue)
as a function of $r$ 
predicted by our models. 
({\bf Bottom}): 
The corresponding radial profile of velocity anisotropy 
$\beta(r) = 1 - (\sigma_\theta^2+\sigma_\phi^2)/(2\sigma_r^2)$. 
In both panels, the coloured dashed lines bracket the central 95 percentile of the posterior distribution of our model. 
The central 68 percentile of $\sigma_r$ and $\beta$ 
are also shown by the grey shaded region. 
We do not show 68\% region for $\sigma_\theta$ and $\sigma_\phi$ for clarity. 
Open symbols are the velocity dispersions and the velocity anisotropy of K giants \citep{Bird2019}, which are not used in our fit but are shown for reference. 
}
\label{fig:veldisp}
\end{figure}

\begin{figure*}
\centering
\includegraphics[width=3.2in]{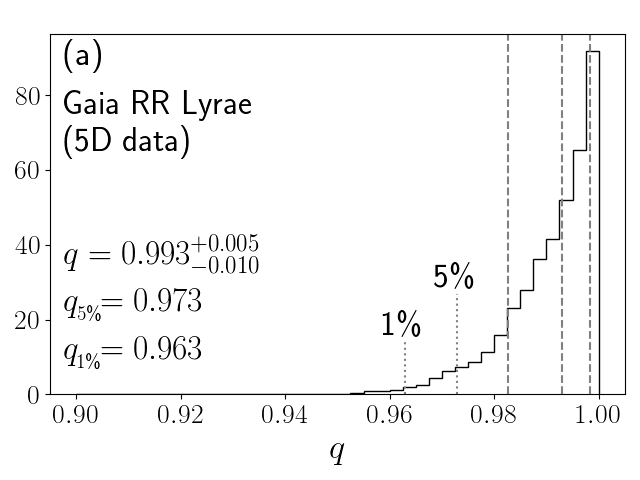}
\includegraphics[width=3.2in]{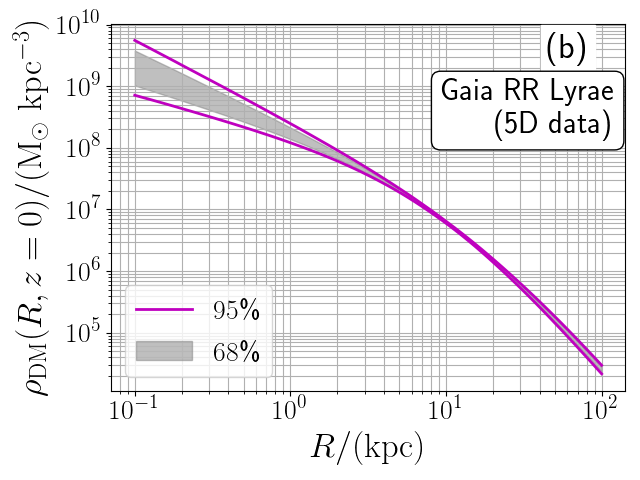}\\
\includegraphics[width=3.2in]{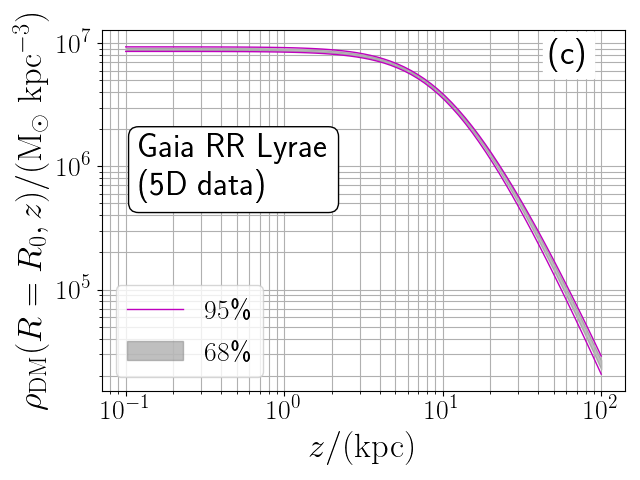}
\includegraphics[width=3.2in]{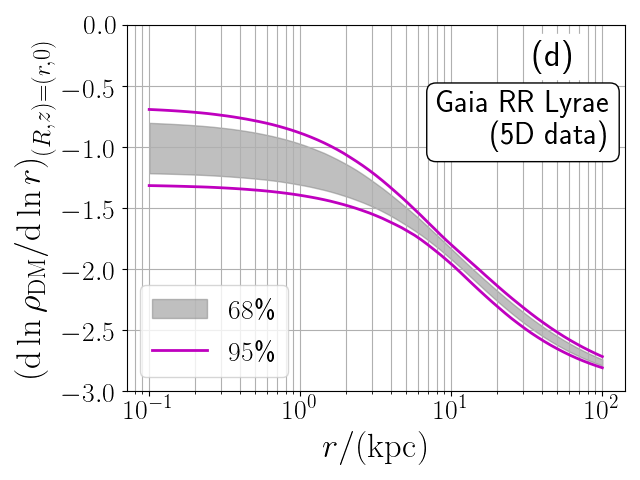}\\
\caption{
The posterior distribution of the dark matter density profile. 
\textbf{(a)}
The probability distribution function of the density flattening $q$. 
The three dashed vertical lines at $q>0.98$ 
corresponds to $(16,50,84)$ percentiles of the distribution. 
The lower 1 and 5 percentiles  
are located at $q=0.963$ and $0.973$, respectively. 
The posterior distribution 
rules out oblate dark halo models with $q<0.963$ 
with a confidence level of $99\%$. 
We note that the parameter range at $q>1$ is not explored in this paper. 
\textbf{(b)} The density profile $\rho_\mathrm{DM}(R,0)$ estimated on the Galactic disc plane as a function of the cylindrical radius $R$. The shaded region corresponds to the central 68\% of the distribution, while the magenta solid lines denote the central 95\% of the distribution. 
\textbf{(c)} The same as panel (b), but for the density profile $\rho_\mathrm{DM}(R=R_0,z)$ estimated at the Solar cylinder as a function of the distance $z$ from the Galactic plane. 
\textbf{(d)} 
The logarithmic density slope 
$\mathrm{d} \ln \rho_\mathrm{DM}/ \mathrm{d} \ln r$ 
evaluated on the Galactic plane $(R,z)=(r,0)$. 
}
\label{fig:GaiaDMdensity}
\end{figure*}


\section{Discussion}
\label{sec:discussion}

\subsection{Comparison with other studies}

We have analysed the kinematics of RR Lyrae stars to estimate the DM density distribution, 
especially focusing on the flattening $q$ of the DM halo within 30~kpc. Our result indicates that $q>0.963$ with 99\% confidence level, which is consistent with a nearly spherical DM halo, although we cannot currently explore the possibility that $q>1$ (prolate). Here, we compare our result with previous studies.

\subsubsection{Previous results with stellar streams}

\cite{Koposov2010} modelled GD-1 stellar stream \citep{GD2006} and estimated the flattening of the Galactic potential.  They found that the DM potential's flattening is $q_{\Phi} > 0.89$ with 90\% confidence level. 
According to a relationship between the 
density flattening and the potential flattening,\footnote{
For a potential model with a nearly flat rotation curve, 
the density flattening $q$ and the potential flattening $q_\Phi$ 
are related by 
$q^2 \simeq 2 q_\Phi^4 - q_\Phi^2$  
(see equation 2.72b of \citealt{BT2008}). 
} 
their constraint corresponds to 
a density flattening of $q > 0.68$ with 90\% confidence level. 
\cite{Bovy2016}  
measured the DM density flattening to be 
$q=1.3^{+0.5}_{-0.3}$ when using GD-1 stream (at $r=14 \kpc$) 
and 
$q=0.93\pm0.16$ when using Pal 5 stream (at $r=19 \kpc$). 
By combining these two data sets, 
they estimated the global value of $q$ to be $q=1.05 \pm 0.14$. 
More recently, 
\cite{Malhan2019} analysed the GD-1 stream 
by using the astrometric data from Gaia DR2 
to estimate 
$q = 0.82^{+0.25}_{-0.13}$. 
Although the statistical uncertainties are relatively large, 
all of the above-mentioned results using the GD-1 stream 
are consistent with a nearly spherical DM halo. 

 \cite{Law2010} modelled the Sagittarius stellar stream to 
conclude that the best-fitting Galactic DM halo model 
is oblate-triaxial with  
the major axis lying in the Galactic disc plane about $7^\circ$ to the Galactocentric $y$-axis, the intermediate axis  perpendicular to the stellar disc, and the short axis lying $7^\circ$ from the Galactic $x$-axis. The DM density distribution is flattened such that the axis lengths of the {\it density distribution} along Galactocentric $x,y,z$ are given by $x:y:z = 0.44:1:0.97$. This model of the halo is strongly disfavoured since simulations show that stellar discs perpendicular to the intermediate axis of such a halo are violently unstable \cite{Debattista2013}. In addition,
\cite{Pearson2015} argued that the triaxial potential model by \cite{Law2010} is not a good approximation at least at $r<20 \kpc$ since it would cause too much dispersal and thickening to the Pal 5 stellar stream.

Recently \cite{Vasiliev2020Tango} constructed a halo model that fits Gaia DR2 proper motion data as well as all available radial velocity data with a time-dependent Galactic halo model that includes the reflex motion resulting from the gravitational perturbation by the Large Magellanic Cloud (LMC). These authors find that the models that fit the Sagittarius stream best, include deformation to the MW DM distribution such that the halo is oblate with an axis ratio $R : z \simeq 1 : 0.6$ and aligned with the disc in the inner part of the halo, but becoming triaxial (twisted and then prolate-triaxial) and misaligned  with the disc beyond $\sim 50 \kpc$.

\subsubsection{Previous results with field halo tracers}

\cite{Loebman2014} applied the axisymmetric Jeans equations 
to kinematic data for field halo stars from SDSS and estimated the DM halo's density flattening to be $q \simeq 0.4 \pm 0.1$. This estimate is significantly smaller than most other studies (expect for the recent work based on the Sagittarius stream when influenced by the LMC; \citealt{Vasiliev2020Tango}). Interestingly, they also found that their halo sample has a radial velocity dispersion $\sigma_r \simeq 141 \kms$ across the survey volume ($d \lesssim 10 \kpc$), while we find $\sigma_r \simeq 180 \kms$ near the Sun dropping to $\sigma_r \simeq 160 \kms$ at $R\sim 20$~kpc. If our RR Lyrae sample and their halo sample trace the same population of halo stars, this disagreement might arise from two sources. First \cite{Loebman2014} used a proper motion sample derived from SDSS and POSS which has significantly larger errors than Gaia DR2 proper motions. Second, they used photometric distances to their field star sample which are less accurate than distances to the Gaia RRLyrae sample. 

\cite{Wegg2019} applied a similar axisymmetric Jeans equation formalism to an RR Lyrae sample from Gaia DR2 and estimated the DM density flattening to be $q \simeq 1.00 \pm 0.09$. Their sample highly overlaps with our RR Lyrae sample, and their spatial selection cut is similar to ours.  The fact that our result is also consistent with a spherical DM halo 
provides strong evidence that the DM halo within $r<30$~kpc is not highly oblate (however see Section~\ref{sec:disequilibrium} and Appendix  \ref{sec:validation_m12m} for the effects of disequilibrium).

It is worth mentioning that \cite{Posti2019} modelled the kinematics of globular clusters 
with an action-based DF model 
and estimated $q=1.3 \pm 0.25$ (prolate). 
They used \agama\ to compute the orbital actions.  
However, the method to compute actions that is implemented in \agama\ is inapplicable to prolate potentials, but the package does not explicitly forbid their use. 
Thus, the users of \agama\ have to determine whether their application is appropriate. 
In this regard, the validity of their analysis is questionable.

\subsubsection{Prediction from numerical simulations}

As mentioned in Section \ref{sec:intro}, numerous  cosmological hydrodynamical simulations over the past 15 years have
predicted that the DM halo of MW-sized galaxies 
have oblate axisymmetric shapes within the inner (0.15-0.3)$r_{200}$. The most recent value of the mean flattening based on several thousand galaxies from the Illustris simulations being  $\langle{q}\rangle = 0.79 \pm 0.15$ \citep{Chua2019} within $\sim 0.15r_{200} \sim 30$~kpc.  This is much flatter than the $q$ value obtained from our analysis, which excludes an oblate halo with $q<0.963$ with a confidence level of 99\%. This could either imply a tension between the predictions of cosmological hydrodynamical simulations and our results or it could imply that some of our assumptions, principally, the assumption of dynamical equilibrium (as discussed in Section \ref{sec:disequilibrium} and Appendix  \ref{sec:validation_m12m}), could be in doubt. At this time, additional applications of this method to mock data from simulations with haloes that are out of equilibrium, e.g. due to the interaction with the LMC, are needed to assess the source of this disagreement.

\subsection{Some issues in our analysis}

\subsubsection{Dynamical disequilibrium}
\label{sec:disequilibrium}

We have assumed that the MW is in dynamical equilibrium. However, this assumption might be too simplistic. For example, \cite{Iorio2019} pointed out that the RR Lyrae sample used here shows a 
triaxial spatial distribution within $r<30$~kpc that has its principal axes tilted relative to the principal axes of the Galactic potential, and is possibly misaligned with the Galactic disc. This triaxial distribution of RR Lyrae stars in  the inner stellar halo is thought to have been deposited by a highly radial accretion event referred to as the ``Gaia-Sausage'' \citep{Belokurov2018} or ``Gaia-Enceladus'' \citep{Helmi2018Nature}. Additional evidence for disequilibrium comes from the observation of two prominent substructures in the RRLyrae sample, the Hercules-Aquila Cloud and the Virgo Overdensity \citep{Simion2019}.

In this regard, it is worth mentioning that we also applied our code to a mock data set generated from one galaxy m12m from the Fire-2 Latte cosmological hydrodynamical suite of simulations \citep{Wetzel2016, Hopkins2018,Sanderson2020}. Like the real MW, this galaxy is not in perfect dynamical equilibrium and includes halo substructure. Our modelling of this galaxy results in an overestimate of the value of $q$ (see Fig. \ref{fig:m12m_DMdensity_validation_q}). This is in contrast with our analysis with mock data sets generated from smooth, equilibrium halo models, which successfully recovers the input values of $q$ with no obvious systematic bias (see Fig.~\ref{fig:DMdensity_validation_q}). 
Thus if the RR Lyrae stars in the inner halo used in our analysis and  \cite{Wegg2019} are significantly out of equilibrium, the assumption of dynamical equilibrium could have resulted in an overestimate of $q$.  While analysis of several more similar cosmological simulations is needed to assess how much of an overestimate can be expected from disequilibrium our analysis of  galaxy m12m  suggests an inflation of $\Delta q \sim 0.1-0.2$ implying that the true flattening could be closer to $q \sim 0.75-0.90$.

\cite{Erkal2020} argued that the perturbation from LMC is not negligible at $r \gtrsim 30 \kpc$. If the LMC's perturbation is strong, our DF fitting method might result in a biased estimate of $q$ (see also \citealt{Petersen2020arXiv}). 
However, the inner part of the halo is less affected by such a perturbation \citep{Garavito2020arXiv}, so our analysis may not be seriously affected by LMC's perturbation. Recently, \cite{Vasiliev2020Tango} modelled the dynamics of the MW, LMC, 
and the Sagittarius dwarf galaxy.  
They also estimated the radial profile of the DM halo's shape based on the morphology of the Sagittarius stream. In principle, it is possible to formulate how LMC affects the DF, 
but this is beyond the scope of this paper (see \citealt{Deason2020arXiv}).

\subsubsection{More general shapes for the dark matter halo}

Throughout this paper, we have assumed that $q$ is constant as a function of radius. However, if $q$ changes as a function of $r$, as predicted by cosmological hydrodynamical simulations \citep[e.g.][]{Zemp2012} our estimate of $q$ might be biased. In principle, we can relax the assumption of constant $q$ with some extra parameters, such as the inner and outer values of $q$ and the transition radius.

We also note that, it is possible to estimate the triaxiality of the DM halo 
by implementing a fast algorithm to compute actions in a general triaxial potentials (including prolate potentials, with long axis oriented perpendicular to the disc plane). 
One possibility is to use the algorithm by \cite{Sanders2015triaxial}. The fact that our posterior distribution of $q$ is peaked at $q=1$, the upper boundary of the currenly explored range of $q$, implies that future investigations of prolate and triaxial shapes should be carried out. 

\subsubsection{Metallicity dependence of the distribution function}
\label{sec:discussion_FeH_dependence}

There is some observational evidence that the DF of the stellar halo depends on the metallicity \citep{Carollo2007, Carollo2010, Deason2011, Hattori2013, Kafle2013, Das2016a, Bird2019, CarolloChiba2020arXiv, Iorio2020arXiv}. 
In this paper, we do not take into account the metallicity dependence, because it would increase the number of free parameters. 
Our sample is confined to the inner halo ($r \lesssim 30 \kpc$), where relatively metal-rich halo stars ([Fe/H]$>-2$) dominate, therefore the DF is probably most representative of metal rich stars. However, if we were to apply our method to a sample of stars in a larger volume (say $r \lesssim 100 \kpc$), the metallicity dependence would be more important.


\section{Conclusions} \label{sec:conclusion}

In this paper, we combined proper-motion data in Gaia DR2 for the halo RR Lyrae stars within $d \leq 20 \kpc$ from the Sun \citep{Iorio2019}, circular velocity data for red giants in the disc plane from Gaia DR2 \citep{Eilers2019}, and the vertical force data from APOGEE \citep{BovyRix2013} to constrain the 3D shape of the Galactic DM halo, by assuming that the stellar halo can be described by 
an analytic DF model and that the MW is axisymmetric. 

Our method is based on the DF fitting formulation that was pioneered by \cite{McMillanBinney2012,McMillanBinney2013}, and elaborated upon by \cite{Ting2013} and \cite{Trick2016}. The most important contribution of our method is to introduce a new way to handle the distance uncertainty of sample stars.  

Our results can be summarised as follows:
\begin{itemize}
\item 
99\% of the posterior distribution of  $q=c/a$ (minor-to-major axis ratio) is located at $q>0.963$ (see Fig.~\ref{fig:GaiaDMdensity}(a)). 
We emphasise that we only explored oblate models with $q \leq 1$ due to a limitations in the way we compute orbital actions of halo stars.

\item 
Our estimated value of  $q>0.963$ implies a nearly spherical DM halo within $r \lesssim 30 \kpc$ and strongly disfavours a very flattened DM halo. This may be in conflict with recent $\Lambda$CDM cosmological simulations that predict $\langle{q}\rangle = 0.79 \pm 0.15$ \citep{Chua2019} within $0.15r_{200} (\sim 30 \kpc)$. 

\item 
While validation tests of our code with with mock data created from smooth, equilibrium galactic models recover the values of $q$ to high accuracy (see Fig.~\ref{fig:DMdensity_validation_q}), our test with mock data generated from a galaxy (m12m) from the Fire-2 Latte cosmological hydrodynamical suite of simulations yields $q$ values overestimated by $\sim 0.1-0.2$ (see Fig.~\ref{fig:m12m_DMdensity_validation_q}). This implies that if the MW halo is not in dynamical equilibrium as we have assumed, our estimate of $q$ is an overestimate. 

\item
Our derived distribution function is a good match to the proper motion distribution in $(l,b)$ and the derived correlation coefficient $\rho_{\muell\mub}$ shows the quadrupole feature characteristic of the radially anisotropic distribution observed in the data (see Fig.~\ref{fig:pmlb}). The derived distribution function also provides a estimates of the radial, azimuthal and polar velocity dispersion profiles ($\sigma_r(r), \sigma_\phi(r), \sigma_\theta(r)$) and velocity anisotropy $\beta(r)$ that are a good match to observed velocity dispersion and anisotropy profiles of K giant stars \citep{Bird2019}, which were not used in our analysis (see Fig.~\ref{fig:veldisp}).

\item 
Our result puts a tight constraint on 
the local DM density: $\rho_{\mathrm{DM},\odot} =$  $0.00901^{+0.00018}_{-0.00020} M_\odot \pc^{-3}$, 
or 
 $0.342^{+0.007}_{-0.007}$ 
$\;\mathrm{GeV} \mathrm{cm}^{-3}$(see Fig.~\ref{fig:GaiaDMdensity}(c) and Table \ref{tab:DM_summary}), which is consistent with other recent estimates \citep{Read2014,BlandHawthorn_Gerhard2016}. 

\item 
Our result favours a cuspy DM halo with inner density slope  
$(-\gamma) = -\left( 0.982^{+0.227}_{-0.197} \right)$, 
which is consistent with an NFW profile (see Fig.~\ref{fig:GaiaDMdensity}(b)(d) and Table~\ref{tab:DM_summary}).

\end{itemize}

\section*{Acknowledgements}

K.H. thanks Giuliano Iorio for kindly sharing the clean sample of RR Lyrae stars in \cite{Iorio2019}, and Robyn Sanderson and Andrew Wetzel for kindly sharing the Latte simulations. 
K.H. thanks Sergey Koposov for useful conversations and for his support. 
K.H. thanks Pablo F. de Salas for frequent discussion that improved K.H.'s analysis code. 
K.H. thanks A.H. for the support during this work. 
K.H. \& M.V. thank members of the stellar halos group at the University of Michigan for continued camaraderie and stimulating discussion. M.V. and K.H. were supported by NASA-ATP award NNX15AK79G. M.V. is also supported by NASA-ATP award 80NSSC20K0509. This work was supported by a Michigan Institute for Computational Discovery and Engineering (MICDE) catalyst grant for FY2019. 
Some part of this research was started at the KITP workshop `Dynamical Models for Stars and Gas in Galaxies in the \Gaia\ Era' held at the Kavli Institute for Theoretical Physics. 
This work has made use of data from the European Space Agency (ESA)
mission {\it Gaia} 
(\url{http://www.cosmos.esa.int/gaia}), 
processed by the {\it Gaia} Data Processing and Analysis Consortium (DPAC,
\url{http://www.cosmos.esa.int/web/gaia/dpac/consortium}). 
Funding for the DPAC has been provided by national institutions, in particular
the institutions participating in the {\it Gaia} Multilateral Agreement.

We used the following packages:
\agama\ \citep{Vasiliev2019}, 
\texttt{constrNMPy} (\url{https://github.com/alexblaessle/constrNMPy}), 
\texttt{corner.py} \citep{ForemanMackey2016},
\texttt{cubature} (\url{https://github.com/stevengj/cubature}), 
\texttt{emcee} \citep{emcee}, 
\texttt{gizmo\_read} (\url{https://bitbucket.org/awetzel/gizmo_read}), 
\texttt{matplotlib} \citep{hunter07},
\texttt{numpy} \citep{vanderwalt11},
\texttt{PyGaia} (\url{https://github.com/agabrown/PyGaia}), 
and 
\texttt{scipy} \citep{jones01}. 



\bibliographystyle{mnras}
\input{mybibtexfile.bbl} 



\appendix

\section{Coordinate system} \label{appendix:coordinate}

We adopt a Galactocentric Cartesian coordinate system $(x,y,z)$, in which the $(x,y)$-plane is the Galactic disc plane. The $x$-axis is directed from the Sun to the Galactic centre, the $y$-axis is parallel to the direction of the Galactic rotation at the Solar position (i.e., the direction of $\ell=90^\circ$ as seen from the Sun), 
and the $z$-axis is perpendicular to the Galactic disc. 
The position of the Sun is assumed to be $\vector{x}_\odot = (x_\odot,y_\odot,z_\odot) = (-R_0,0,0)$, with $R_0 = 8.178 \kpc$ \citep{Gravity2019}. The velocity of the Sun with respect to the Galactic rest frame is assumed to be 
$\vector{v}_\odot = (v_{x,\odot},v_{y,\odot},v_{z,\odot}) = (11.10, 247.30, 7.25) \kms$ 
\citep{Reid2004, Schonrich2010}. 
We also define a Galactocentric spherical coordinate system $(r,\phi,\theta)$ and a Galactocentric cylindrical coordinate system $(R,\phi,z)$
, 
such that $(x,y,z)=(r \cos\theta \cos\phi, r \cos\theta \sin\phi, r \sin\theta) = (R \cos\phi, R \sin\phi, z)$. 
Also, for each 3D location with respect to the Sun, we define the line-of-sight unit vector $\vector{e}_\mathrm{los}$.


\section{Full result of our MCMC analysis} \label{appendix:corner}

\begin{figure*}
\centering
\includegraphics[width=6.8in]{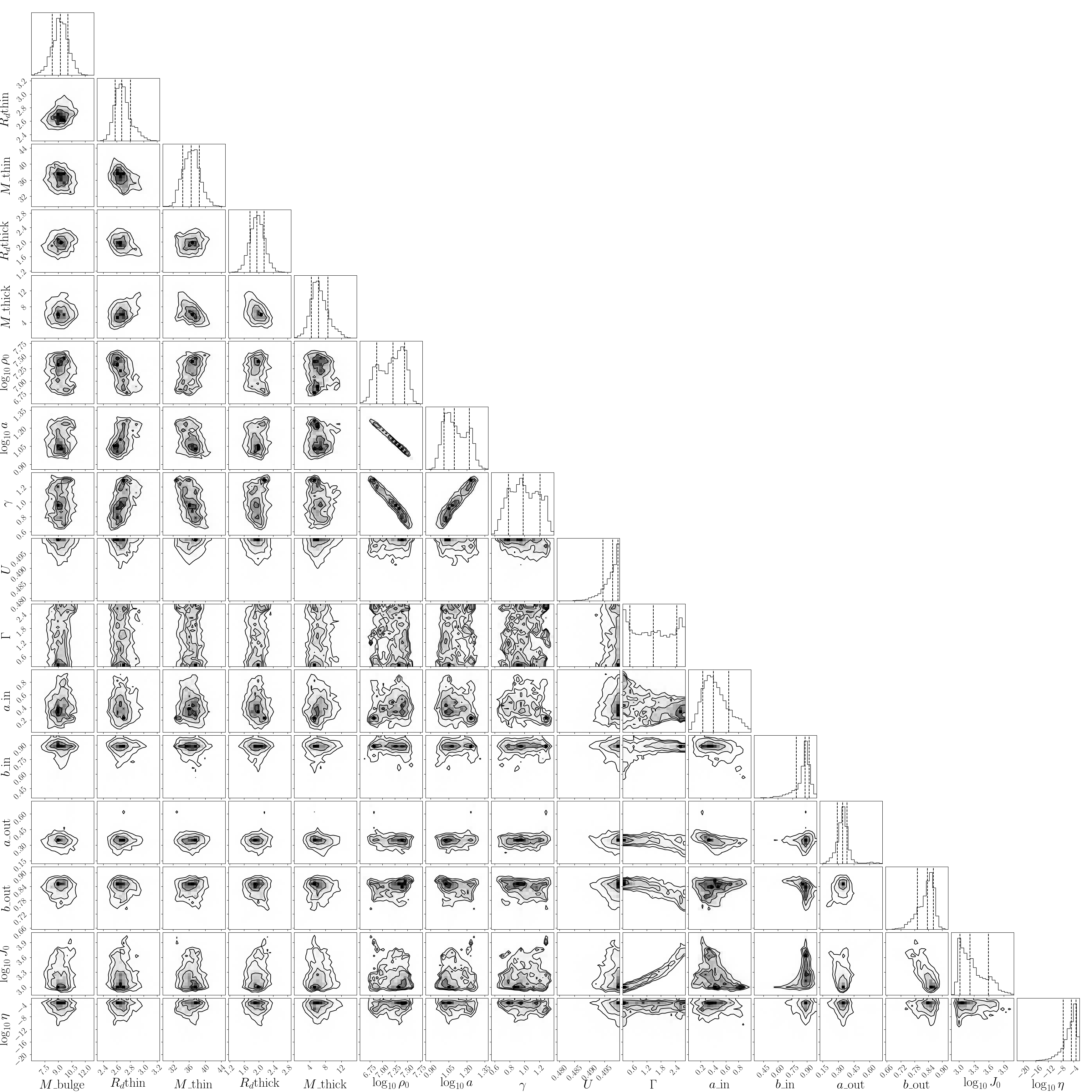}
\caption{
The corner plot of all the parameters in our analysis 
of Gaia RR Lyrae stars. 
The parameters shown here are 
the parameters for the baryonic mass distribution 
$(M_\mathrm{bulge}/(10^{10} M_\odot)$, 
$R_\mathrm{d}^\mathrm{thin}$, 
$M_\mathrm{thin}/(10^{10} M_\odot)$,
$R_\mathrm{d}^\mathrm{thick}$, 
$M_\mathrm{thick}/(10^{10} M_\odot))$, 
the parameters for the dark matter mass distribution 
$(\log_{10} \rho_0 $,
$\log_{10} a$,
$\gamma$, 
$U)$,
and the parameters for the distribution function 
$(\Gamma$, $a_\mathrm{in}$, 
$b_\mathrm{in}$, 
$a_\mathrm{out}$, 
$b_\mathrm{out}$,
$\log_{10} J_0$,  
$\log_{10} \eta)$. 
%
%
}
\label{fig:corner_raw_all}
\end{figure*}

\begin{figure*}
\centering
\includegraphics[width=6.8in]{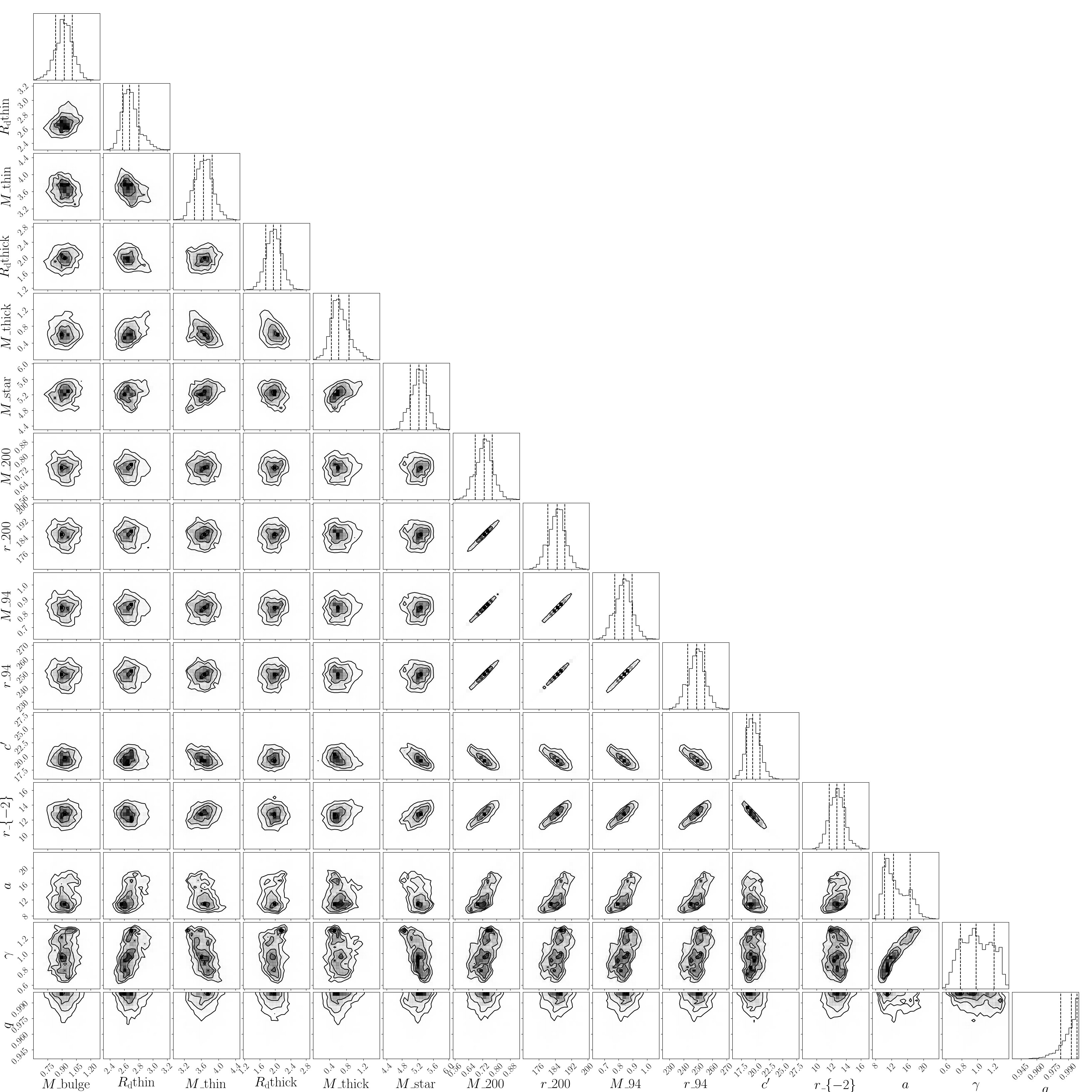}
\caption{
The same as Fig. \ref{fig:corner_raw_all}, 
but expressed by more physically meaningful quantities. 
The parameters shown here are 
the parameters for the baryonic mass distribution 
$(M_\mathrm{bulge}/(10^{10} M_\odot)$, 
$R_\mathrm{d}^\mathrm{thin}$, 
$M_\mathrm{thin}/(10^{10} M_\odot)$,
$R_\mathrm{d}^\mathrm{thick}$, 
$M_\mathrm{thick}/(10^{10} M_\odot)$, 
$M_\mathrm{star}/(10^{10} M_\odot))$ 
and 
the parameters for the dark matter mass distribution 
$(M_{200}/(10^{12} M_\odot)$, 
$r_{200}$, 
$M_{94}/(10^{12} M_\odot)$, 
$r_{94}$, 
$c'$, 
$r_{-2}$, 
$a$, 
$\gamma$, 
$q)$. 
}
\label{fig:corner_alt_all}
\end{figure*}

Figures \ref{fig:corner_raw_all} and \ref{fig:corner_alt_all} 
show the corner plot 
of the posterior distribution of our main analysis. 
The quantities shown in Fig. \ref{fig:corner_raw_all} 
are the raw variables used in our MCMC, 
while the quantities shown in Fig. \ref{fig:corner_alt_all} 
are more physically meaningful quantities.

\section{Validation of our method with mock data}
\label{sec:validation}

\newcommand{\tableThreeCaption}{
Summary of  mock data used for validation and results. 
}
\newcommand{\tableThreeNote}{
}
\begin{table*}
\centering
\caption{ {\tableThreeCaption} }
\label{tab:mock_summary}
\begin{tabular}{lll lll} 
\hline
Base model & 
True DM density flattening & 
Data type & 
$\rho_\mathrm{DM}(R,0,0)$ & 
$\rho_\mathrm{DM}(R_0,0,z)$ & 
$q$\\
\hline
\hline 
Analytic DF model $+$ analytic potential model& 
$q_\mathrm{true} = 0.6$ & 
5D & 
Fig. \ref{fig:DMdensity_validation_R_z0}(a)  & 
Fig. \ref{fig:DMdensity_validation_R0_z}(a)  & 
Fig. \ref{fig:DMdensity_validation_q}(a) \\
Analytic DF model $+$ analytic potential model & 
$q_\mathrm{true} = 0.8$ & 
5D & 
Fig. \ref{fig:DMdensity_validation_R_z0}(b)  & 
Fig. \ref{fig:DMdensity_validation_R0_z}(b)  & 
Fig. \ref{fig:DMdensity_validation_q}(b) \\
Analytic DF model $+$ analytic potential model & 
$q_\mathrm{true} = 0.996$ & 
5D & 
Fig. \ref{fig:DMdensity_validation_R_z0}(c)  & 
Fig. \ref{fig:DMdensity_validation_R0_z}(c)  & 
Fig. \ref{fig:DMdensity_validation_q}(c) \\
Analytic DF model $+$ analytic potential model & 
$q_\mathrm{true} = 0.6$ & 
6D & 
Fig. \ref{fig:DMdensity_validation_R_z0}(d)  & 
Fig. \ref{fig:DMdensity_validation_R0_z}(d)  & 
Fig. \ref{fig:DMdensity_validation_q}(d) \\
Analytic DF model $+$ analytic potential model & 
$q_\mathrm{true} = 0.8$ & 
6D & 
Fig. \ref{fig:DMdensity_validation_R_z0}(e)  & 
Fig. \ref{fig:DMdensity_validation_R0_z}(e)  & 
Fig. \ref{fig:DMdensity_validation_q}(e) \\
Analytic DF model $+$ analytic potential model & 
$q_\mathrm{true} = 0.996$ & 
6D & 
Fig. \ref{fig:DMdensity_validation_R_z0}(f)  & 
Fig. \ref{fig:DMdensity_validation_R0_z}(f)  & 
Fig. \ref{fig:DMdensity_validation_q}(f) \\
\hline 
m12m galaxy & 
Fig. \ref{fig:m12m_axis_ratio}  & 
5D & 
Fig. \ref{fig:m12m_DMdensity_validation_R_z0}(a)  & 
Fig. \ref{fig:m12m_DMdensity_validation_R0_z}(a)  & 
Fig. \ref{fig:m12m_DMdensity_validation_q}(a) \\
m12m galaxy & 
Fig. \ref{fig:m12m_axis_ratio}  & 
6D & 
Fig. \ref{fig:m12m_DMdensity_validation_R_z0}(b)  & 
Fig. \ref{fig:m12m_DMdensity_validation_R0_z}(b)  & 
Fig. \ref{fig:m12m_DMdensity_validation_q}(b) \\
\hline 
\end{tabular} \\
\flushleft{\tableThreeNote}
\end{table*}


\begin{figure*}
\centering
\includegraphics[width=2.2in]{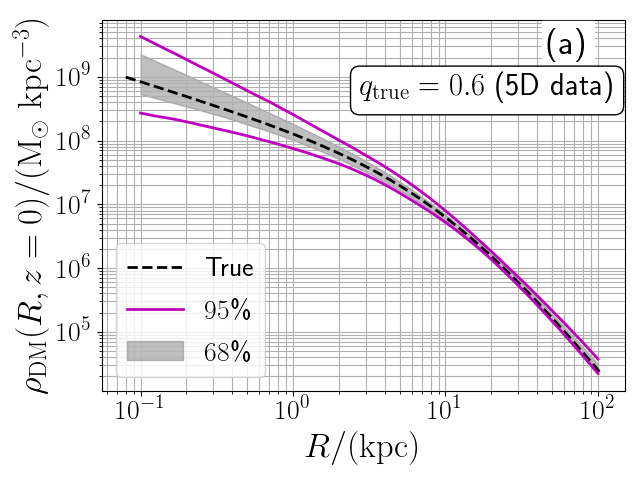}
\includegraphics[width=2.2in]{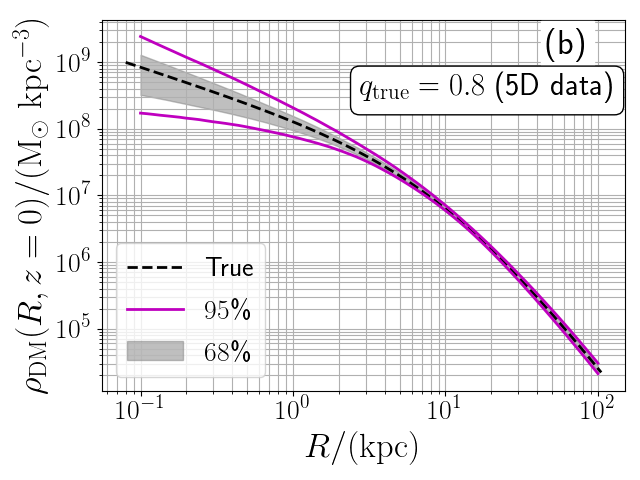}
\includegraphics[width=2.2in]{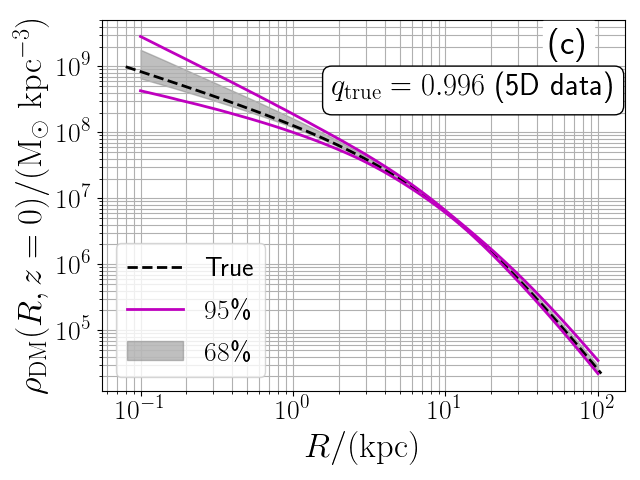}\\
\includegraphics[width=2.2in]{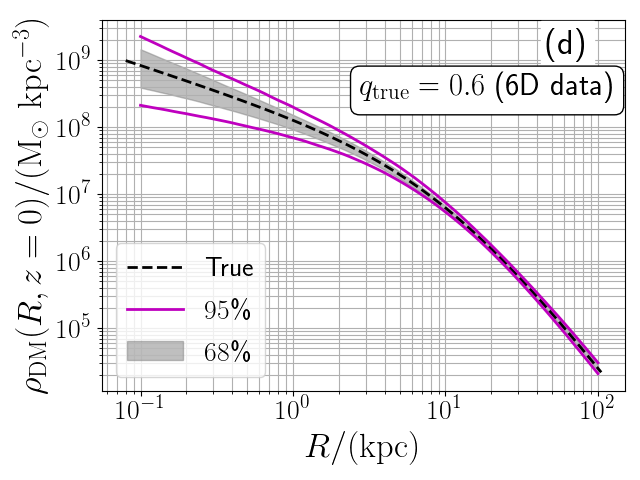}
\includegraphics[width=2.2in]{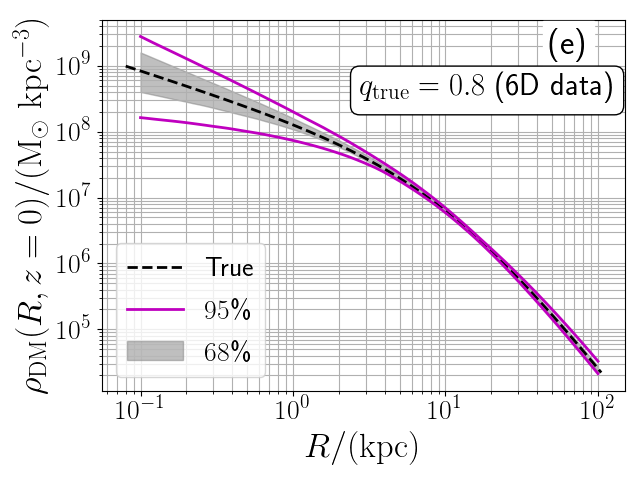}
\includegraphics[width=2.2in]{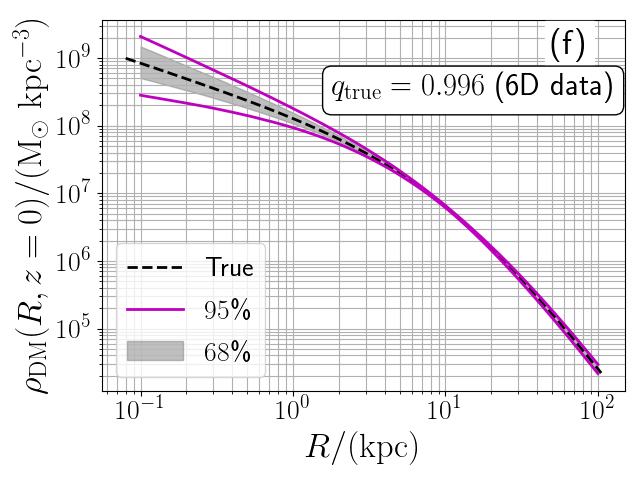}
\caption{
The dark matter density profile $\rho(R,z=0 \kpc)$ reconstructed from our mock analysis. The input value of the flattening, $q_\mathrm{true}$, is shown in each panel. 
The shaded region and the region enclosed by solid lines 
cover the central 68 and 95 percentiles of the posterior distribution. 
Also, the dashed line 
corresponds to the true profile 
of the input model. 
In panels (a)-(c), the results are shown for 
DF fitting with 5D data (without $\vlos$ data). 
In panels (d)-(f), the results are shown for 
DF fitting with 6D data. 
The agreement between the true profile and the posterior profile 
indicates that our method can recover the dark matter profile 
even if we lack the $\vlos$ data. 
}
\label{fig:DMdensity_validation_R_z0}
\end{figure*}

\begin{figure*}
\centering
\includegraphics[width=2.2in]{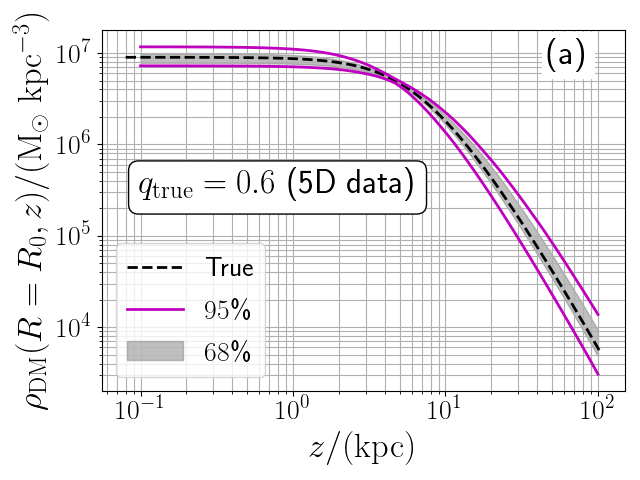}
\includegraphics[width=2.2in]{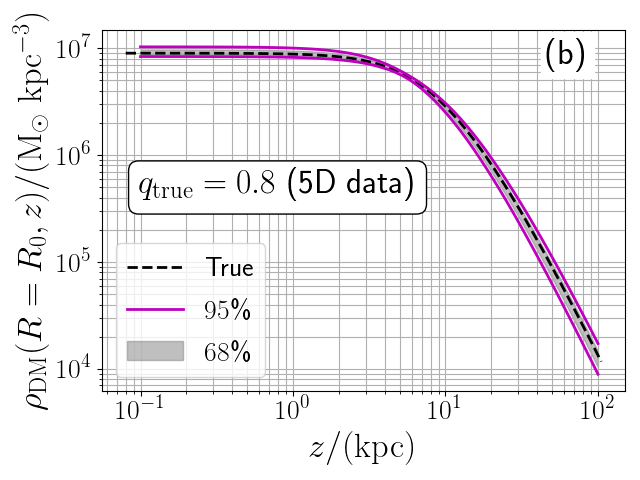}
\includegraphics[width=2.2in]{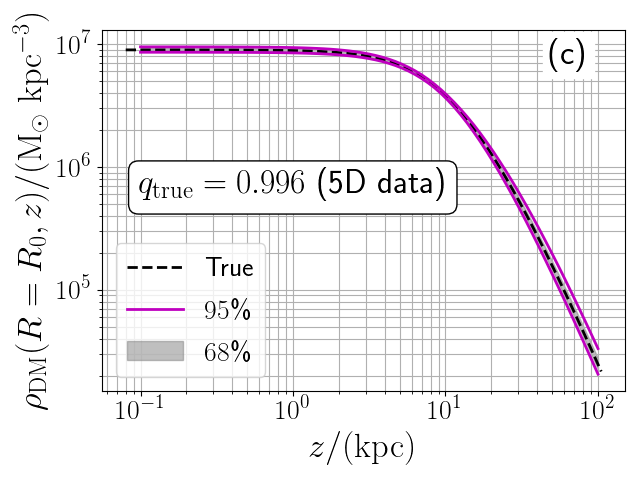}\\
\includegraphics[width=2.2in]{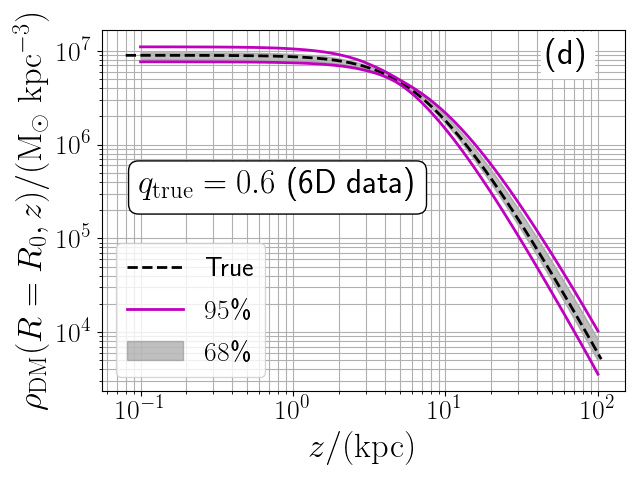}
\includegraphics[width=2.2in]{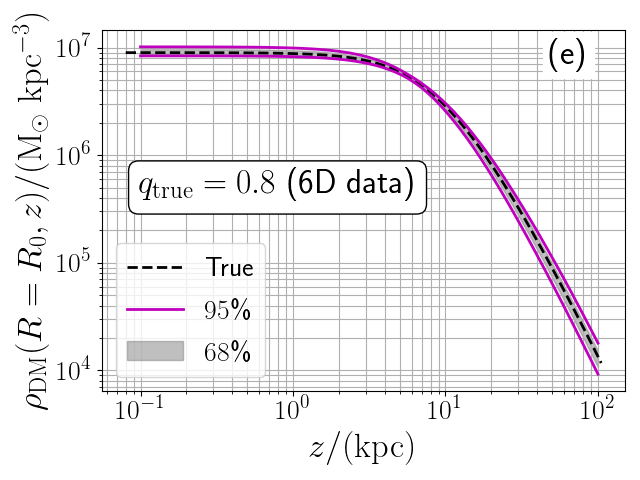}
\includegraphics[width=2.2in]{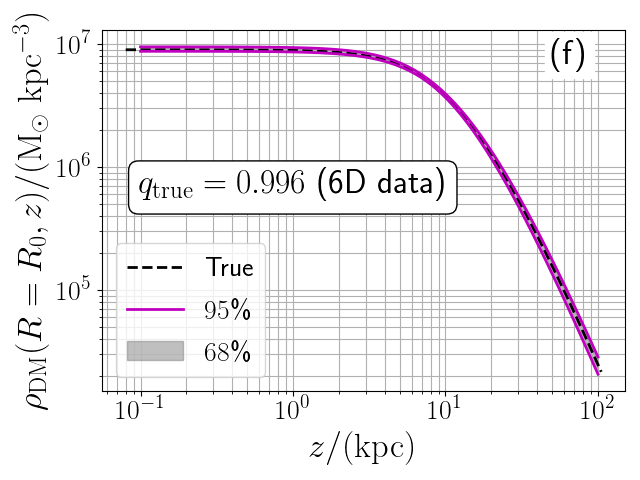}
\caption{
The same as Fig.\ref{fig:DMdensity_validation_R_z0}
but for the reconstructed dark matter density profile 
$\rho(R_0,z)$ 
of the mock data. 
}
\label{fig:DMdensity_validation_R0_z}
\end{figure*}

\begin{figure*}
\centering
\includegraphics[width=2.2in]{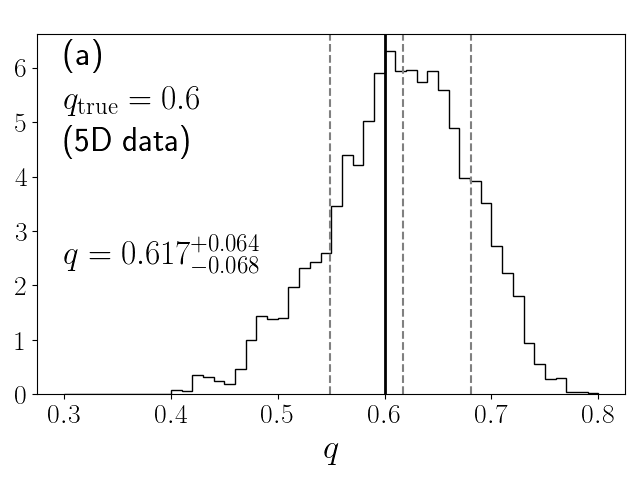}
\includegraphics[width=2.2in]{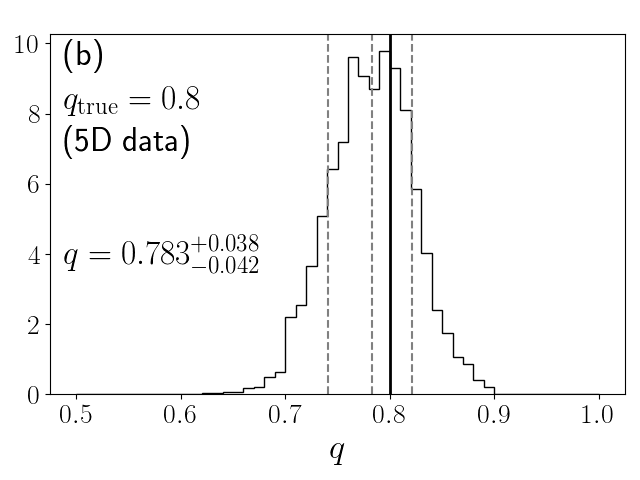}
\includegraphics[width=2.2in]{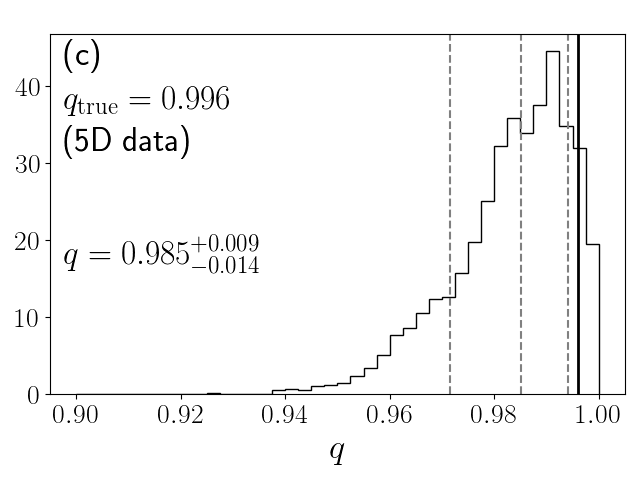}\\
\includegraphics[width=2.2in]{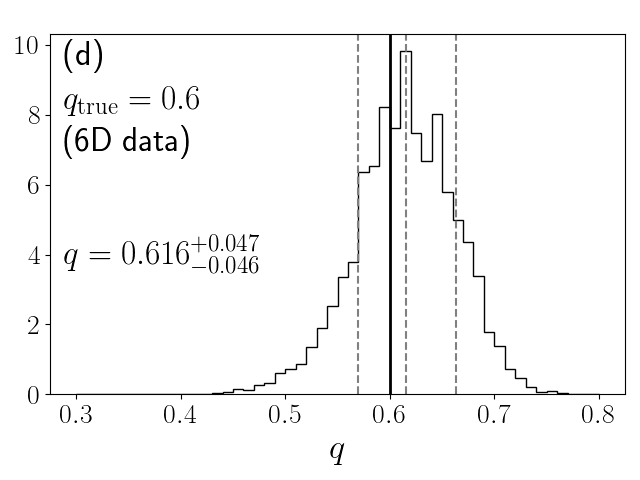} 
\includegraphics[width=2.2in]{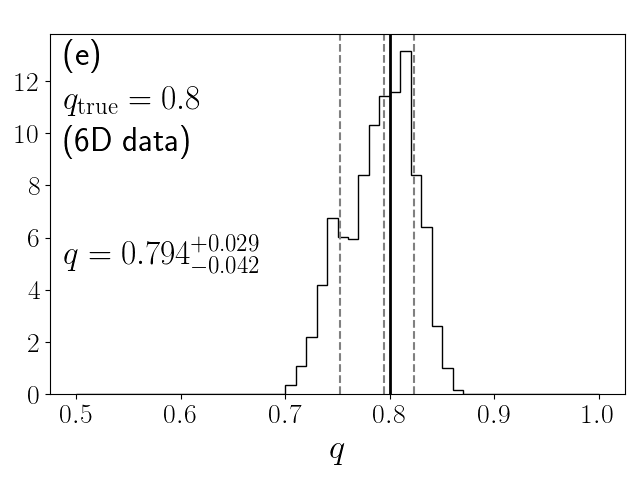} 
\includegraphics[width=2.2in]{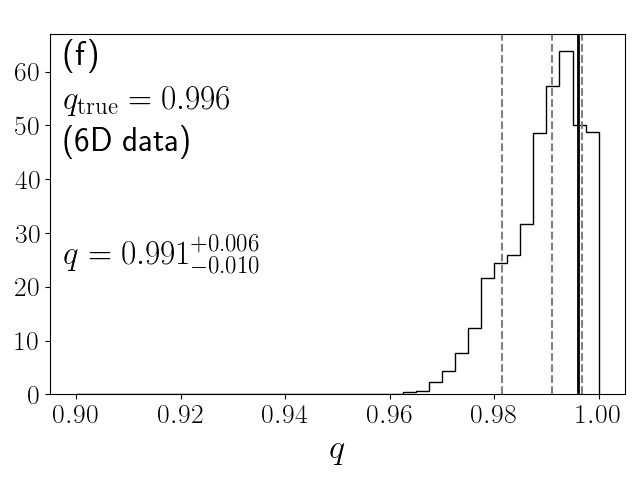} 
\caption{
The posterior distribution of the dark matter density flattening $q$ 
for the mock data. 
We note that the range of horizontal axis is different in each panel. 
}
\label{fig:DMdensity_validation_q}
\end{figure*}

To validate our method and results for the Gaia RR Lyrae sample, 
we perform similar analyses with eight mock data sets 
summarised in Table \ref{tab:mock_summary}. For each of the models below, we generate a 5D mock data set without $\vlos$ and a 6D mock data set with full information with the number and spatial distribution of stars similar to the RR Lyrae sample.

To generate these mock data, we prepare three smooth stellar halo models constructed from analytic DFs embedded in analytic potential models, as detailed in Appendix \ref{sec:smooth_mock}. 
We also generated a mock date set from a realistic Milky-Way-like galaxy m12m from the Fire-2, Latte cosmological hydrodynamical simulation suite \citep{Wetzel2016, Hopkins2018, Sanderson2020}, as detailed in Appendix \ref{sec:m12m_mock}.

\subsection{Validation with `smooth halo' mock data sets} 

\subsubsection{Mock data for smooth stellar halo models} \label{sec:smooth_mock}

We construct three smooth stellar halo models. 
We assume that the halo stars are test particle moving 
in an analytic potential composed of baryonic component and DM component. 
The functional form of the model potential (bulge, thin/thick/gas discs and DM halo) and the functional form of the stellar halo DF are identical to those in Section \ref{sec:model}. 
We adopt the same gas disc model as in \cite{McMillan2017}, 
and we adopt three sets of parameters 
for bulge, thin disc, thick disc, DM halo, and stellar halo DF model:
\begin{itemize}
\item the best-fitting parameters shown in Table \ref{tab:maxL} (with $q=0.996$); 
\item
the same as the best-fitting parameters but with $q=0.8$; and
\item
the same as the best-fitting parameters but with $q=0.6$. 
\end{itemize}
In other words, these models are different from each other only in terms of the DM density flattening $q$. 

For each model, we evaluate the circular velocity $v_\mathrm{circ}^\mathrm{model}(R)$ at the same radii as in the data of \cite{Eilers2019} and take into account the same amount of observational error. 
Also, for each model, we evaluate the vertical force $K_{z,1.1 \kpc}^\mathrm{model}(R)$ at the same locations as in the data of \cite{BovyRix2013} and take into account the same amount of fractional error. 
The mock data of circular velocity and vertical force  are used to aid our DF fitting.

To generate mock RR Lyrae stars, we first sample mock halo stars from the above-mentioned three input models. We then add Gaia DR2-like proper motion errors that depend on the Gaia's $G$-band photometry. For this purpose, we assume that all the mock stars are RR Lyrae stars 
with $G$-band absolute magnitude\footnote{
We assume that RR Lyrae stars have a colour $(V-I)=0.6$ and $V$-band absolute magnitude $M_V=0.6$.
We derive $(G-V)$ and 
$M_G = M_V + (G-V)$
with \texttt{PyGaia} (\url{https://github.com/agabrown/PyGaia}).
} 
$M_G = -0.1376$, 
and compute Gaia DR2-like proper motion error by using a formula described in equation (16) in \cite{Sanderson2020}. 
We also add 
a distance modulus uncertainty of 0.240 (mimicking RR Lyrae stars), and an optimistic $5 \kms$ error on $\vlos$. 

From this sample of error-added mock stars, 
we randomly choose $N=16197$ stars by using the spatial selection function defined in Section \ref{sec:selection_function}. In the following, 
we use three sets of `6D mock data' and 
three sets of `5D mock data' for which we mask $\vlos$.

\subsubsection{Radial dark matter density}

Fig. \ref{fig:DMdensity_validation_R_z0} shows 
the reconstructed DM density 
$\rho_\mathrm{DM}(R, z=0 \kpc)$. 
We see that the central 68 percentile region (grey shaded region) traces the true DM profile (dashed line) for all mock data.

Although the spatial distribution of our mock sample stars is limited by the survey volume, 
the inferred density profile traces 
the actual density profile even beyond the survey volume. Given that the inner density slope is allowed to freely vary, it is intriguing that 
the DF fitting can recover well 
the radial DM density profile 
at $1<R/\kpc<5$. 
The almost perfect reconstruction of the outer density profile at $R\gtrsim 30 \kpc$ 
can be partly explained by the fact that 
we fixed the outer density slope to be the correct value ($-3$).

\subsubsection{Vertical dark matter density}

Fig. \ref{fig:DMdensity_validation_R0_z} shows 
the reconstructed vertical profile of the DM density 
$\rho_\mathrm{DM}(R=R_0, z)$ 
evaluated at the Solar cylinder. 
We see that the true profile (dashed line) 
matches well with the central 68 percentile region (grey shaded region) 
for all the mock data sets. 
By comparing panels (a)-(c) and (d)-(f), 
we see that the uncertainty in $\rho_\mathrm{DM}(R=R_0, z)$ is larger for models with more flattened DM halo (with smaller $q_\mathrm{true}$). This trend may be explained 
by the fact that a more flattened DM distribution makes it harder to disentangle the dynamical contribution from DM and the baryonic discs.

\subsubsection{Dark matter density flattening}

Fig. \ref{fig:DMdensity_validation_q} shows the posterior distribution of $q$. We see that the posterior distribution is centred around $q_\mathrm{true}$, for both 5D and 6D mock data. This result implies 
that the lack of $\vlos$ is not a problem in inferring $q$ as long as we handle the missing $\vlos$ properly.

It is not surprising that the uncertainty in $q$ is larger if we use 5D data than we use 6D data. 
This result indicates that the added information  from $\vlos$ improves the inference on $q$, 
which motivates us to obtain $\vlos$ for a large number of Galactic RR Lyrae stars.

We note that the uncertainty in $q$ is smaller when $q_\mathrm{true}$ is larger. For example, for the 5D mock data sets, the difference between the 16th and 50th percentiles of the posterior distribution is 
$\Delta q = 0.068$, $0.042$, and $0.014$ for $q_\mathrm{true}=0.6, 0.8$, and $0.996$, respectively. 
This trend may be understood by considering that it is more difficult to distinguish the dynamical contributions from the baryonic disc and the DM halo 
if the DM halo is more flattened.

In this paper, we do not allow $q>1$, because \agama\ can compute the actions only in oblate potentials. In Fig. \ref{fig:DMdensity_validation_q}(c)(f), where $q_\mathrm{true}=0.996$, the posterior distribution is skewed because our method cannot explore the parameter space at $q>1$. 
It is worth noting that 
the posterior distribution of $q$ is peaked at $0.99<q<1$, 
consistent with $q_\mathrm{true}=0.996$. 
This result is intriguing 
when we interpret our results of Gaia RR Lyrae sample, 
where the posterior distribution is peaked at $q=1$, 
(see Fig. \ref{fig:GaiaDMdensity}(a)). 
Based on our mock analysis, 
our results with Gaia RR Lyrae stars might indicate that 
the Galactic DM halo is very close to spherical. 

\subsection{Validation with realistic mock data generated from the m12m galaxy from the Latte simulations} 
\label{sec:validation_m12m}

\subsubsection{True shape of the dark matter halo of galaxy m12m} \label{sec:m12m_axis_ratio}

We first derive the true shape of the m12m's DM halo   
with an iterative `S1 method' in \cite{Zemp2011}. 
Fig. \ref{fig:m12m_axis_ratio} shows the minor-to-major axis ratio $(c/a)$ and the intermediate-to-major axis ratio $(b/a)$ of the DM density at each ellipsoidal radius $m$ defined by 
\eq{ \label{eq:def_m}
m^2  = x_a^2 + \left(\frac{x_b}{b/a}\right)^2 + \left(\frac{x_c}{c/a}\right)^2 . 
}
Here, $(a,b,c)$ ($a \geq b \geq c > 0$) are the length scale of the three principal axes of the DM density and $(x_a, x_b, x_c)$ are the spatial coordinates along $(a,b,c)$-axes, respectively. It turns out that the minor axis ($c$-axis) of m12m DM halo is approximately parallel to the $z$-axis within $m<100 \kpc$. 
At $m<8 \kpc$, we see that $(b/a, c/a) \simeq (0.8, 0.65)$, 
so the DM halo is triaxial. At $m>12 \kpc$, in contrast, we see that $(b/a, c/a) \simeq (0.95, 0.65)$, which means that the DM halo become nearly oblate-axisymmetric with $q=(c/a)=0.65$ beyond $\sim 12$kpc.

\subsubsection{Mock data generated from LATTE simulation} \label{sec:m12m_mock}

By using the star, gas, and DM particles in m12m galaxy, we evaluate the circular velocity curve and the radial profile of the vertical force at $z=1.1 \kpc$ above the disc plane. We use these quantities to generate mock data of circular velocity curves and vertical force profiles akin to those shown in Fig.~\ref{fig:circ} and Fig.~\ref{fig:Kz}  used in modelling the real RR Lyrae data.

To generate mock samples mimicking the RR Lyrae stars, we first select old halo stars in m12m  with metallicity [Fe/H]$<-1.5$ and age $\tau> 8 \Gyr$. For these old halo stars, we add mock observational error 
assuming that all the stars are RR Lyrae stars. As in Section \ref{sec:smooth_mock}, 
we prepare a 6D mock data set 
and a 5D mock data set without $\vlos$ data whose spatial distribution is defined in Section \ref{sec:selection_function}.

We use these mock halo catalogues along with 
the mock circular velocity curve and vertical force, to infer the DM density distribution in m12m in exactly the same manner as done for the Gaia RR Lyrae sample in the main body of this paper.

\subsubsection{Radial dark matter density of m12m galaxy} 

Fig. \ref{fig:m12m_DMdensity_validation_R_z0} 
shows the posterior distribution 
of the DM density profile at $(R,z)=(R,0 \kpc)$. 
We see that $\rho_\mathrm{DM}(R, 0)$ is   
successfully reconstructed for $1 \lesssim R \lesssim 100 \kpc$. Given that we do not use mock halo stars within $5 \kpc$ from the galactic centre, the successful recovery of the inner density profile of the DM halo 
is quite promising. Our tests with m12m mock data suggest that the density profile of the Galactic DM halo 
shown in Fig. \ref{fig:GaiaDMdensity}(a) 
is reliably recovered at $1 \kpc \lesssim R$.

It seems to be challenging for our method to recover 
the DM density profile within $1 \kpc$ from the galaxy. 
For example, Fig. \ref{fig:m12m_DMdensity_validation_R_z0} shows that 
the reconstructed profile has a steeper density slope than the true profile  
at $R \lesssim 1 \kpc$. 
This is not surprising, 
as we have seen in Fig. \ref{fig:DMdensity_validation_R_z0} 
that the reconstructed density profile at $R\lesssim 1 \kpc$ 
is associated with large uncertainty 
even if 
the mock data are generated from a smooth halo model. 
This problem might be resolved by future access to halo stars within $5 \kpc$ from the galactic centre, however this is observationally challenging.

\subsubsection{Vertical dark matter density of m12m galaxy} 

Fig. \ref{fig:m12m_DMdensity_validation_R0_z} shows the posterior distribution 
of the DM density profile at $(R,z)=(R_0,z)$. 
We see that the global shape of $\rho_\mathrm{DM}(R_0, z)$ is 
more or less well recovered, 
although there is some offset 
such that the reconstructed profile shows a lower density at low-$|z|$  and the `knee' of the recovered density profile occurs at slightly larger $|z|$ than it should.

\subsubsection{Dark matter density flattening of m12m galaxy} 

Fig. \ref{fig:m12m_DMdensity_validation_q} shows the posterior distribution of $q$. The three vertical dashed lines shows the (16,50,84)th percentiles of the distribution. We see that the posterior distribution has a width of $\sim 0.1$ around its peak, which is comparable to that in Fig. \ref{fig:DMdensity_validation_q}(a)(d).

\begin{figure}
\centering
\includegraphics[width=3.in]{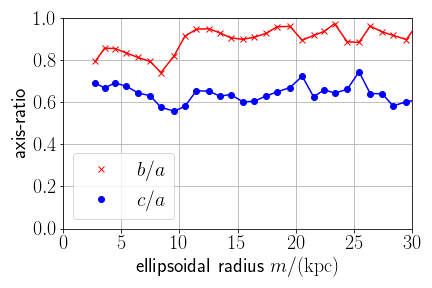}
\caption{
The minor-to-major axis ratio $(c/a)$ 
and the intermediate-to-major axis ratio $(b/a)$ 
as a function of the ellipsoidal radius $m$ from the 
centre of m12m galaxy. 
We see that 
the inner part ($m<8\kpc$) of the dark matter halo is triaxial, 
while it becomes nearly axisymmetric at $m>12 \kpc$. 
}
\label{fig:m12m_axis_ratio}
\end{figure}

\begin{figure}
\centering
\includegraphics[width=2.2in]{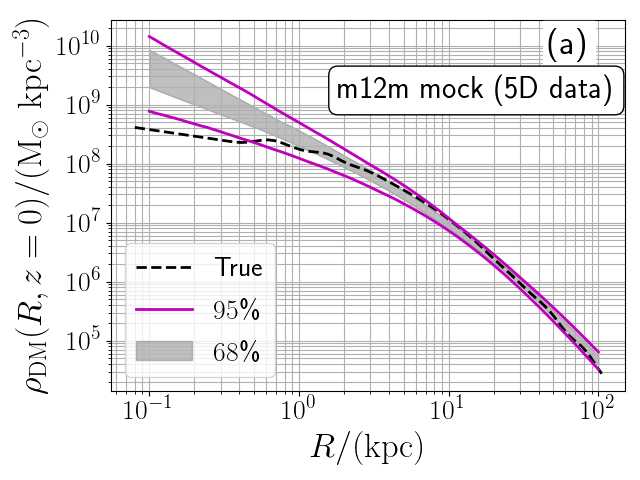}\\
\includegraphics[width=2.2in]{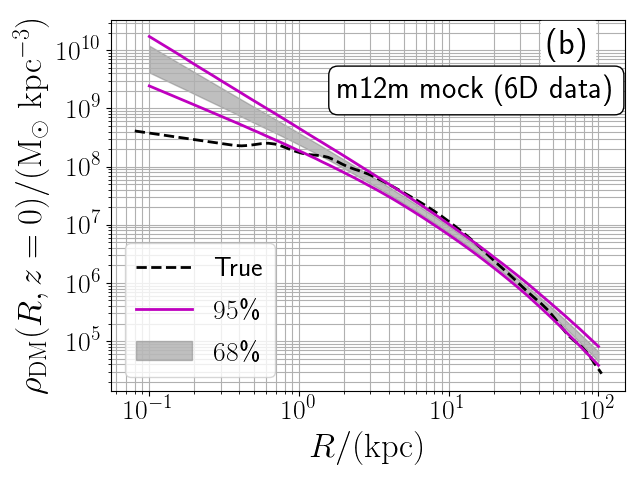}
\caption{
The same as Fig. \ref{fig:DMdensity_validation_R_z0} 
but using 5D and 6D mock data sets generated from m12m galaxy. 
The dashed line shows the true density profile $\rho_\mathrm{DM}(R, z=0 \kpc)$, 
which is estimated from 
spherical harmonic expansion of the dark matter particles in m12m galaxy. 
}
\label{fig:m12m_DMdensity_validation_R_z0}
\end{figure}

\begin{figure}
\centering
\includegraphics[width=2.2in]{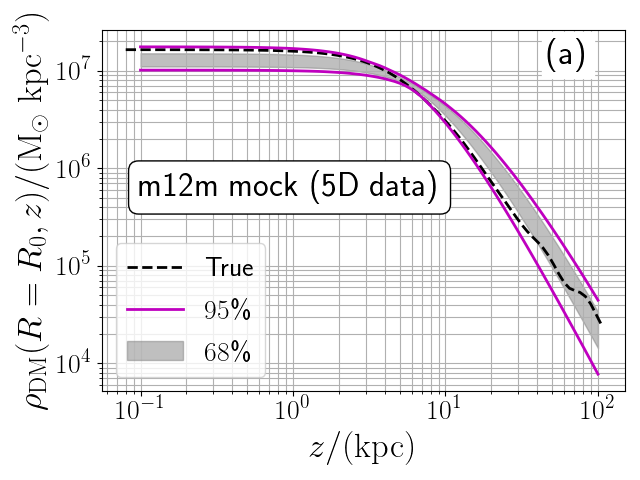}\\
\includegraphics[width=2.2in]{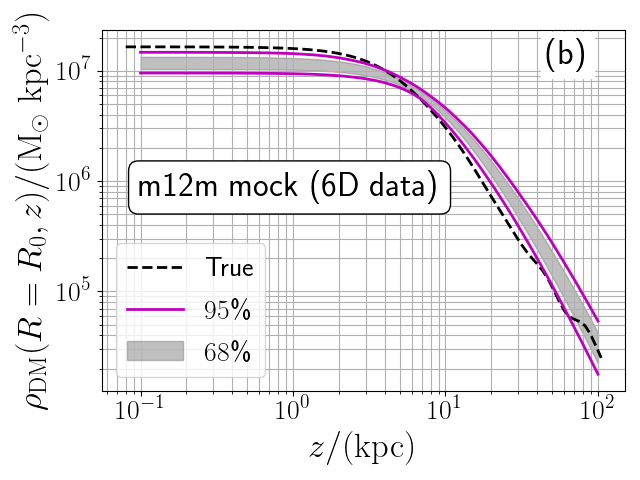}
\caption{
The same as Fig. \ref{fig:DMdensity_validation_R0_z} 
but using 5D and 6D mock data sets generated from m12m galaxy in LATTE simulation. 
The dashed line shows the true density profile $\rho_\mathrm{DM}(R_0, z)$, 
which is estimated from 
spherical harmonic expansion of the dark matter particles in m12m galaxy. 
}
\label{fig:m12m_DMdensity_validation_R0_z}
\end{figure}

\begin{figure}
\centering
\includegraphics[width=2.2in]{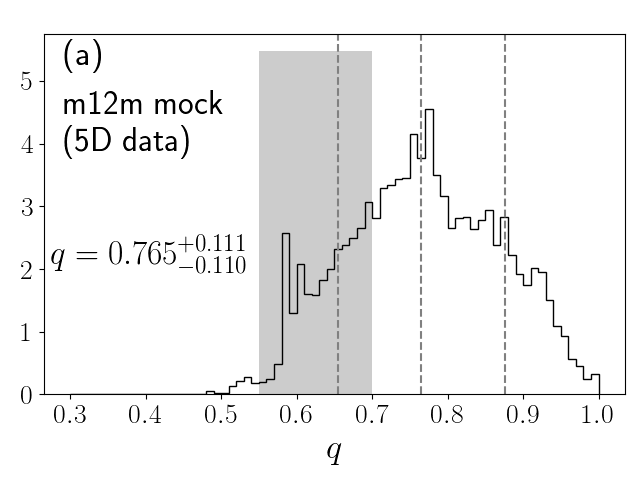}\\
\includegraphics[width=2.2in]{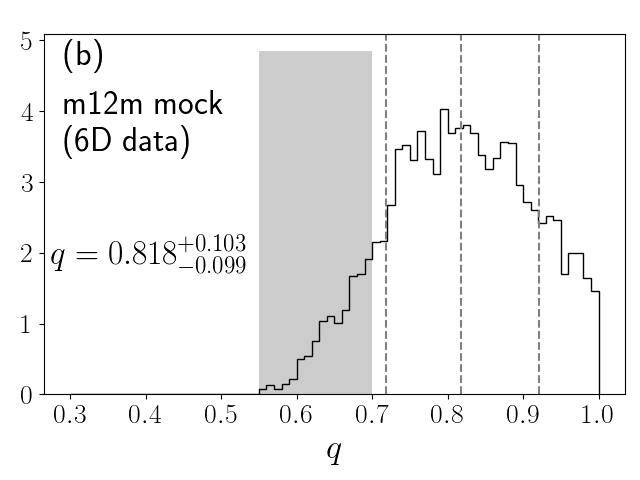}
\caption{
The same as Fig. \ref{fig:DMdensity_validation_q} 
but using 5D and 6D mock data sets generated from m12m galaxy. 
The shaded vertical region at $0.55 < q < 0.7$ 
corresponds to the true 
density flattening as inferred from 
the radial profile of $(c/a)$ in Fig. \ref{fig:m12m_axis_ratio}. 
}
\label{fig:m12m_DMdensity_validation_q}
\end{figure}

\begin{figure}
\centering
\includegraphics[width=3.in]{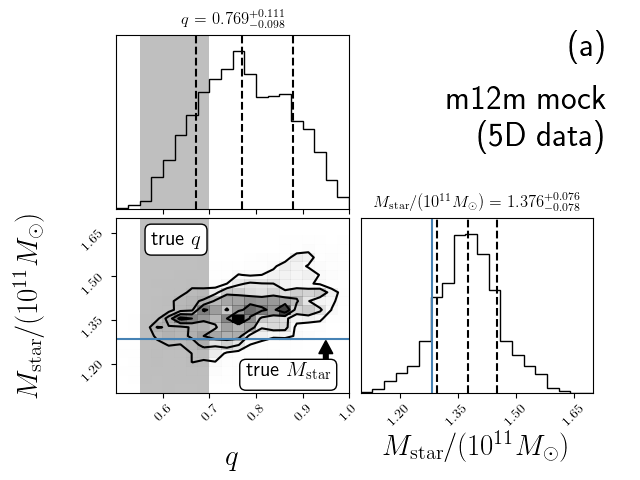}\\
\includegraphics[width=3.in]{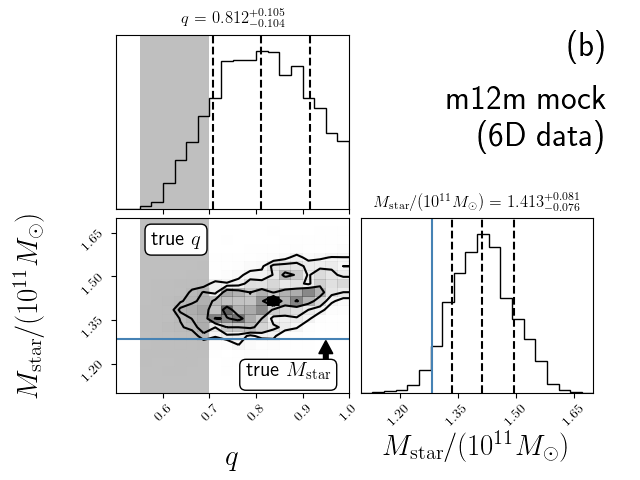}
\caption{
The correlation between the total stellar mass $M_\mathrm{star}$ and 
the dark matter density flattening $q$ 
seen in the posterior distribution of the mock analysis of m12m galaxy. 
The correct value of $q$ is 
$0.55<q<0.7$, 
which is shown by a grey vertical band 
in the left-hand panels. 
The correct value of $M_\mathrm{star}$ is 
$M_\mathrm{star} = 1.28\times10^{11}M_\odot$,   
which is shown as the horizontal solid line in lower left corner 
and as the vertical solid line in the lower right corner 
of each panel. 
}
\label{fig:m12m_DMdensity_validation_q_vs_Mbaryon}
\end{figure}

The true flattening of m12m DM halo is $0.55<q<0.7$ (see the axis ratio $(c/a)$ in Fig. \ref{fig:m12m_axis_ratio}). 
Thus, the median of the posterior distributions seen in Fig. \ref{fig:DMdensity_validation_q}
is systematically larger than it should be by $\sim 0.1-0.2$. 
To understand the origin of this systematic offset, 
we show in Fig. \ref{fig:m12m_DMdensity_validation_q_vs_Mbaryon} 
the correlation between the 
total stellar mass $M_\mathrm{star}$ 
and the DM flattening $q$. 
The positive correlation between $(q, M_\mathrm{star})$ 
can be intuitively understood:
If $M_\mathrm{star}$ is overestimated, 
then the overall potential will become more flattened 
unless the DM halo becomes `less flattened' to compensate. 
Thus, in our Bayesian analysis, 
the increase of $M_\mathrm{star}$ is balanced by the increase of $q$. 
We speculate that our model favours a parameter set 
such that the stellar mass is larger than it should be, 
and that results in an overestimation of $q$. 
We have confirmed that 
our estimate of $q$ in m12m's DM halo can be improved (but not perfectly) if we fix the baryonic potential to the ground-truth potential computed from the particle distribution in m12m galaxy. At this moment, it is unclear why large $(q, M_\mathrm{star})$ is favoured in our analysis. 
It might be due to the disequilibrium of the halo, 
or due to the slight triaxiality of the DM halo (see Fig. \ref{fig:m12m_axis_ratio}).  
In any case, the lesson learned from this analysis is that it is important to have strong constraints on the baryonic mass distribution in order to accurately estimate $q$.

\section{A note on Gaia EDR3}

In this paper, we have used the proper motion data from Gaia DR2. 
Just before we complete this work, Gaia EDR3 was released \citep{GaiaEDR3_Brown2020arXiv}. 
We found that the uncertainty in proper motion data for our RR Lyrae star sample is reduced by a factor of $\sim2$, almost independent of their mean $G$-magnitude. We think the main conclusion of our paper will not be changed if we use Gaia EDR3 data instead of Gaia DR2 data, because the dominant sources of uncertainty in our analysis is the uncertainty in the photometric distance of RR Lyrae stars and the lack of $\vlos$ data, both of which will not be improved by Gaia EDR3.

\bsp	
\label{lastpage}
\end{document}